\documentclass{aastex631}

\begin{document}

\title{A Low-mass Model of The Milky Way: The Disk Warp Resulting from A Galaxy Merger}

\author[0000-0002-0592-7660]{Mingji Deng}
\affiliation{School of Astronomy and Space Sciences, University of Chinese Academy of Sciences, Beijing 100049, P.R. China}

\author[:0000-0002-3954-617X]{Cuihua Du}
\affiliation{School of Astronomy and Space Sciences, University of Chinese Academy of Sciences, Beijing 100049, P.R. China}

\author{Jian Zhang}
\affiliation{School of Astronomy and Space Sciences, University of Chinese Academy of Sciences, Beijing 100049, P.R. China}

\author{HaoYang Liu}
\affiliation{School of Astronomy and Space Sciences, University of Chinese Academy of Sciences, Beijing 100049, P.R. China}

\author{ZhongCheng Li}
\affiliation{School of Astronomy and Space Sciences, University of Chinese Academy of Sciences, Beijing 100049, P.R. China}

\correspondingauthor{Cuihua Du}

\email{ducuihua@ucas.ac.cn}



\begin{abstract}

 Previous studies have shown that disk warps can result from galaxy mergers. Recent research indicates a noticeable decline in the rotation curve (RC) of the Milky Way (MW), suggesting the need for a new low-mass model to describe its dynamical features. This study constructs a new Gaia-Sausage-Enceladus (GSE) merger model to characterize the RC features of our galaxy. We use the GIZMO code to simulate mergers with various orbital parameters to investigate how the disk warp evolves under different conditions. This simulation demonstrates the evolutionary mechanism of disk warp, which arises due to the asymmetric gravitational potential of the dark matter (DM) halo generated universally by galaxy mergers. The results indicate that the tilt angle of the DM halo partly reflects the gravitational strength at the $Z=0$ plane, while the gravitational strength on the disk plane reflects the amplitude of disk warp. We identify a dual-regime interaction mechanism driven by the asymmetric halo potential. On short timescales, we find a distinct anti-correlation between the halo's tilt angle and the disk's warp amplitude, indicating a `seesaw' mechanism of angular momentum exchange. On secular timescales, however, dynamical friction drives a global alignment, causing both the halo tilt and the warp amplitude to decay simultaneously. Furthermore, we demonstrate that high-inclination mergers can sustain long-lived prograde precession, where the persistent yet decaying gravitational torque maintains the prograde bending mode against differential wind-up.    

\end{abstract}

\keywords{Galaxy structure (622); Milky Way disk (1050); Milky Way dark matter halo (1049); Galaxy mergers (608); Hydrodynamical simulations (767)}


\section{Introduction} \label{sec:intro}

   Disk warps are a common feature of galaxies, which is documented in previous studies \citep{SS90, SS03, Reshetnikov}. The Milky Way (MW), a typical example of disk and spiral galaxy, also exhibits a clear disk warp, confirmed by various works \citep{Kerr, Levine, Freudenreich, Chen, Shen}. Numerous mechanisms of the Galactic warp have been proposed, including: inflow of intergalactic matter into the halo \citep{OB, QB, JB}; direct gas accretion onto the Galactic disk \citep{LC, Roskar, GS}; intergalactic magnetic fields \citep{BJ}; interactions between satellites such as Sagittarius dwarf galaxy or Magellanic Clouds \citep{Bailin, WB, Laportea, Laporteb, Bennett, Poggio21} with the disk; bending instabilities that generate self‑excited or internally driven warps \citep{HT, NT, RP,BT, SD}; and misalignment of the dark matter halo with the disk \citep{Han23}. Understanding the warp’s formation and evolution is crucial for reconstructing the dynamical history of the MW. 

   Precise measurements based on mono-age stellar samples have revealed the detailed characteristics of the Galactic warp, and several studies have proposed that it originates from external perturbations \citep[e.g.,][]{Poggio, Cheng}. However, these works generally favor recent interactions with satellite galaxies that produce transient warps, which contradict the long-standing persistence of the Galactic warp \citep{Roskar, LC14, Li23}. Moreover, \citet{Bosma} reported that at least half of spiral galaxies exhibit warps, implying the existence of a long-lived and universal mechanism. Within the framework of $\Lambda$CDM cosmology, a defining feature is hierarchical assembly \citep[e.g.,][]{White}, suggesting that warps may form universally through galaxy mergers \citep{Hammer}.

   By combining the long-standing feature of the MW’s S-shaped warp, which has persisted for over 5 Gyr, with its timeline that may coincide with the completion of the gas-rich Gaia-Sausage-Enceladus (GSE) merger \citep{Ciuca}, we previously developed a Galactic warp evolution model primarily driven by the GSE merger \citep{Deng}. This model offers insights into the long-term evolution of the Galactic disk warp. Substantial evidence indicates that the GSE, regarded as the MW’s last major merger, constitutes the bulk of the inner halo \citep[e.g.,][]{Belokurov, Helmi, Naidu2020} and has consequently been extensively studied through analog simulations \citep[e.g.,][]{Bignone, Elias, Fattahi, Grand, Helmi, Koppelman}. In \citet{Deng}, our merger model builds upon the tailored GSE merger simulations of \citet{Naidu2022}, which employed a grid of 500 idealized merger runs with the GADGET code \citep{Springel05, Springel21} to identify a fiducial model that best matches the H3 survey data \citep{Conroy}. By incorporating a gas component into this framework, we successfully reproduced the Galactic disk warp.

   With the release of Gaia DR3 \citep{G3}, several new studies have applied previously developed techniques to infer distances for stars at large radii (even out to 30 kpc) and to fit the $v_c(R)$ rotation curve of MW \citep[e.g.][]{Jiao, Ou}. Their results show a sharp decrease in the Galactic rotation curve, two estimated total mass of the Milky Way is very similar and the average value would be $\mathrm{M_{tot}=2.06^{+0.24}_{-0.13}\times10^{11}M_{\odot}}$, a significant low value, which hints at the need for a new model to reconstruct the dynamical features of the MW.

   Here, we further analyze the mechanism by which galaxy mergers induce warps in the Galactic disk. In this study, we identify the vertical bending mode and investigate its behavior using Fourier decomposition. To examine the connection between the dark matter halo and the disk, we calculate the vertical acceleration exerted by dark matter particles on the disk and evaluate its misalignment relative to the disk plane.
    
   In this study, we deepen our understanding by creating a gas‑rich GSE merger simulation within a low‑mass model to explore how the warp evolves and what causes the disk warp resulting from a galaxy merger. In Section~\ref{sec:Model and simulation}, we provide a description of the simulation model and the initial conditions. In Section~\ref{sec: DMH Simulation}, we illustrate the DM halo profile. In Section~\ref{sec: Gv w}, we show the warp evolution in different orbital settings and how DM halo raises the disk warp. Discussions and results are given in Section~\ref{sec:Discussion} and Section~\ref{conclusion}.

\section{Model and simulation} \label{sec:Model and simulation}

   \begin{table}
    \caption{Initial Conditions at $z\sim2$}\label{ICs}
    $$
      \centering    
         \begin{tabular}{llll}
            \toprule
            \noalign{\smallskip}
            Progenitor Parameters & Milky Way & GSE  & Units\\
            \noalign{\smallskip}
            \hline
            \noalign{\smallskip}
            Dark Matter mass  &  11.55   &  6.93  &  $\mathrm{1\times 10^{10}\,M_{\odot}}$ \\
            $\alpha$          &  3.5     &  3.25  &  N/A \\
            Dark Matter scale radius               &  12.867  &  8.02  &  $\text{kpc}$\\
            Stellar disk mass  &  15    &  0.2   &  $\mathrm{1\times 10^{9}\,M_{\odot}}$\\
            Stellar scale length            &  2     &  0.6   &  $\text{kpc}$\\
            Stellar scale height            &  0.8   &  0.3  &  $\text{kpc}$\\
            Gas disk mass      &  30    &  2.5  &  $\mathrm{1\times10^{9}\,M_{\odot}}$\\
            Gas scale length                &  6     &  2.4  & $\text{kpc}$\\
            Gas scale height                &  3     &  1.2  & $\text{kpc}$\\
            Bulge mass      &  19.5   &  N/A  & $\mathrm{1\times10^{9}\,M_{\odot}}$\\
            Bulge scale                     &  0.9   &  N/A  & $\text{kpc}$\\
            \noalign{\smallskip}
            \hline
            \noalign{\smallskip}
            Orbital Parameters\\
            \noalign{\smallskip}
            \hline
            \noalign{\smallskip}
            Eccentricity ($e$)                   &  0.9  \\
            Initial distance                     &  100.5  & & kpc\\
            Inclination angle                    &  $15^{\circ}, 30^{\circ}, 45^{\circ}, 55^{\circ}, 60^{\circ}, 75^{\circ}$ &   & degree\\
            Relative velocity                    & 73.02 & &km/s\\  
            \noalign{\smallskip}
            \hline
         \end{tabular}
    $$
   \end{table}


   In our previous work \citep{Deng}, we developed a realistic merger model for the MW. This model successfully reproduced the Galactic disk warp, capturing almost all its observed features. However, recent studies \citep{Jiao, Ou} suggest that the MW rotation curve declines more rapidly at larger radii. Therefore, an improved model is needed to accurately represent the mass distribution of the galaxy. The numerical initial conditions (ICs) were generated using the Disk Initial Conditions Environment (DICE) code \citep{Perret14, Perret16}. Density profiles for DM, stars, and gas were input into DICE as distribution functions. These functions were then used to generate Lagrangian particles via the Metropolis-Hastings Monte Carlo Markov Chain algorithm \citep{Metropolis}. The mass resolution is $\mathrm{1\times 10^{5}\,M_{\odot}}$ for baryonic particles and DM particles, such resolution is sufficient to track the warp structures, as the presence of warp has also been observed in some lower resolution cosmological simulation such as TNG100 \citep{Semczuk}. The modeling approach in this study utilizes the GIZMO code \citep{Hopkins15}. We implemented a star formation and feedback module following \citet{Wangj}, enabling gas particles to convert into stellar particles and vice versa through feedback mechanisms. GIZMO is a variant of the smooth particle hydrodynamics code Gadget-3, employing Adaptive Gravity Softening \citep[see details in][]{Hopkins18}.

   \subsection{GSE} \label{sec:GSE}

   Our starting point is based on GSE mass measured by \citet{Lane}. Using high-purity samples of APOGEE DR16 red giants selected by kinematics and chemistry, they estimated the mass of the GSE remnant as $\mathrm{1.45_{-0.51}^{+0.92}\times10^{8}\,M_{\odot}}$, with a total mass $\sim$ $\mathrm{6\times10^{10}\,M_{\odot}}$. The GSE model consists of a stellar disk, a gas disk, and a DM halo. Considering mass loss during the merger process, the total mass of the GSE progenitor used in the simulation is set to $\mathrm{7.2\times10^{10}\,M_{\odot}}$, which is 20\% higher than the reference value \citep{Lee}. The stellar and gas disk models follow an exponential + sech-z profile. The stellar mass is set to the median reference value of $\mathrm{2\times10^{8}\,M_{\odot}}$. The stellar disk scale length $R_{\star}$  (approximately 0.6 kpc), is determined using the size-mass relation (SMR) from \citet{Mowla}. The disk scale height is $\mathrm{0.5\times R_{\star}}$, based on simulations showing that $z\sim2$ disks become thick due to turbulent gas and remain thick \citep{Park}. Given that it is a gas-rich merger at $z\sim2$, we assume a high gas fraction of 0.92 for the GSE. Following assumptions from galaxy merger and metallicity gradient simulations by \citet{Cox} and \citet{Rocha}, the gas disk scale length is set to $\mathrm{4\times R_{\star}}$, with a scale height equal to half of this length. 

   \subsection{Milky Way} \label{sec:Milky Way}
   The MW mass is determined based on the hypothesis that the GSE merger was a major merger with a mass ratio of 2.5 : 1 \citep{Belokurov, Naidu2022}. Therefore, the virial mass of the MW is set to be $\mathrm{1.8\times10^{11}\,M_{\odot}}$. Additionally, we require an appropriate representation of the MW during the merger epoch. Ages derived from various methods suggest that nearly all of the current high-$\alpha$/thick disk and the bulge formed at $z > 1$. Therefore, we model the MW at $z\sim2$ as a combination of the present-day thick disk, gas disk, and bulge. 
   
   \citet{Xiang} studied the age-dependent structure and star formation rate of the MW disk using high-$\alpha$ stars with substantial orbital angular momentum and precise age determinations. Referring to Fig. 4 in their work, we set the stellar disk mass at approximately $\mathrm{1.5\times10^{10}\,M_{\odot}}$ at $z\sim2$. Using the size-mass relation (SMR), the disk scale length ($R_{\star}$) is set to be 2 kpc. The disk thickness is set to $0.4\times R_{\star}$, based on Fig. 3 in \citet{Xiang}. The gas fraction of the MW progenitor is set at 0.667. The scale length of the gas disk is $\mathrm{3\times R_{\star}}$, with a thickness of half scale length. Both disks follow an exponential + sech-z profile. For the MW bulge at $z\sim2$, we adopt the Bulge-E model from \citet{Jiao}, with a mass of $\mathrm{1.95\times10^{10}\,M_{\odot}}$ and a scale length ($R_{b}$) of 0.9 kpc. The bulge is modeled using an Einasto profile \citep{Einasto}, analogous to the Sérsic profile \citep{Sersic} but applicable to three-dimensional mass density distributions \citep{Coe}. This choice is reasonable given that the Sérsic profile is widely used to fit galaxy surface densities. Combining the above components, the total mass is set at $\mathrm{6.72\times10^{10}\,M_{\odot}}$. This value aligns with the baryonic mass of approximately $\mathrm{6\times10^{10}\,M_{\odot}}$ reported by \citet{Jiao} and \citet{Ou}. 

   \subsection{Dark Matter Profile} \label{sec:DM H}

   To accurately reproduce the observational features of the MW rotation curve, we adopted the Einasto profile \citep{Einasto} for the DM halos within the GSE and MW. This profile effectively describes a faster decline observed at large radii \citep{Jiao, Ou}. The Einasto profile is defined as follows                    
   
    \begin{equation}
      \label{f1}
      \rho_{Ein}(r) = \frac{M_{0}}{4\pi r_{s}^{3}}\,exp[-(r/r_{s})^{\alpha}],\\
    \end{equation}

    Here, $M_{0}$ represents the mass normalization, $r_{s}$ is the scale radius, and the parameter $\alpha$ determines the rate at which density decreases with galactic radius—a larger $\alpha$ indicates a faster decline. The parameter $\alpha$ has been measured as approximately 2.3 by \citet{Jiao} and 0.91 by \citet{Ou}. Prior research provides limited guidance regarding suitable $\alpha$ values at $z\sim2$, highlighting a significant gap in the current methodological framework. Enhanced mass deposition in the galactic outskirts due to major mergers naturally flattens the decline of the rotation curve, imposing strict constraints on the initial mass concentrations of progenitor galaxies. After extensive testing, we determined $\alpha$ parameters of 3.5 for the MW progenitor and 3.25 for the GSE progenitor. Corresponding scale lengths ($R_{d}$) were calculated to be $\mathrm{12.867\,kpc}$  and $\mathrm{8.02\,kpc}$, respectively.

    \subsection{Orbital Parameters} \label{sec:OP}
    
    For the orbital configuration, two galaxies with opposite disk spins were placed on a radially-biased, retrograde orbit. The initial separation was determined by summing the virial radii of the two progenitors, while the initial relative velocity was computed based on a Keplerian orbit. Here, we adopted a high eccentricity ($e=0.9$). Such a high value is necessary to reproduce the observed velocity and anisotropy features observed in GSE debris within this low-mass model. Additionally, highly eccentric orbits result in increased star formation rates and enhanced mass accumulation in the galactic inner regions. The inclination angle is another crucial parameter in GSE simulations \citep{Naidu2022, Belokurov23}. In this study, we aim to investigate how mergers influence vertical disturbances in galactic disks. Therefore, we varied the inclination angle while keeping all other parameters constant, setting angles from $15^{\circ}$ to $75^{\circ}$ with intervals of $15^{\circ}$. In our previous work, we modeled the inclination angle as $55^{\circ}$, this angle is included again here. All inclination angles considered in this simulation are: $15^{\circ}$, $30^{\circ}$, $45^{\circ}$, $55^{\circ}$, $60^{\circ}$, and $75^{\circ}$.

    In the DICE configuration, five parameters define the orbit: the polar angle ($\theta$) and azimuthal angle ($\varphi$) of the orbital plane's normal vector, and the spin angles ($\theta_{1},\,\theta_{2},\,\kappa$). Here, $\theta_{1}$ and $\theta_{2}$ represent the angles between the spin vectors of each galaxy and the orbital plane, respectively, while $\kappa$ denotes the angle between the two spin vectors \citep[more detail in Fig. 2 of][]{Perret14}. For example, to position two galaxies on a retrograde orbit with opposite spins and a $15^{\circ}$ inclination, we set the parameters as $(\theta,\, \varphi,\,\theta_{1},\,\theta_{2},\,\kappa)=(0^{\circ}, 165^{\circ}, 165^{\circ}, 15^{\circ}, 180^{\circ})$. After a gas-rich merger, gas particles inherit angular momentum from the orbital momentum of the merger and subsequently redistribute into a thin disk \citep{Barnes}. The subsequent virialization phase, during which the disk rebuilds, typically lasts several gigayears \citep{Hopkins09, Hammer}. During this period, instabilities may induce oscillations that we aim to investigate. Table~\ref{ICs} summarizes the initial conditions used in our simulation.

\section{DM Halo in simulation} \label{sec: DMH Simulation}

   \begin{figure*}
   \label{F1}
   \centering
   \includegraphics[scale = 0.5]{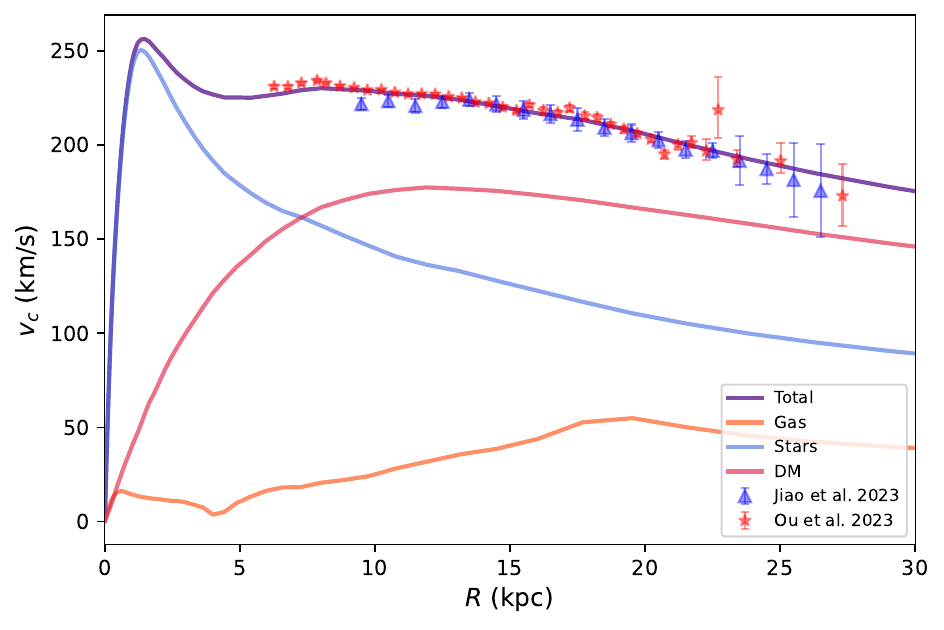}
   \caption{Rotation curve decomposition of the simulated galaxy at $t=11$ Gyr (for the $15^{\circ}$ inclination model). The solid lines represent the circular velocity contributions from different galactic components (DM, Gas, Stars and total), calculated using the PROFILE function of the PYNBODY library \citep{Pontzen}. The data points with error bars indicate observational constraints from the Milky Way, taken from \citet{Jiao} (blue triangles) and \citet{Ou} (red stars). }
              \label{F1}%
    \end{figure*}

   \begin{figure*}
   \label{F2}
   \centering
      \includegraphics[scale = 0.55]{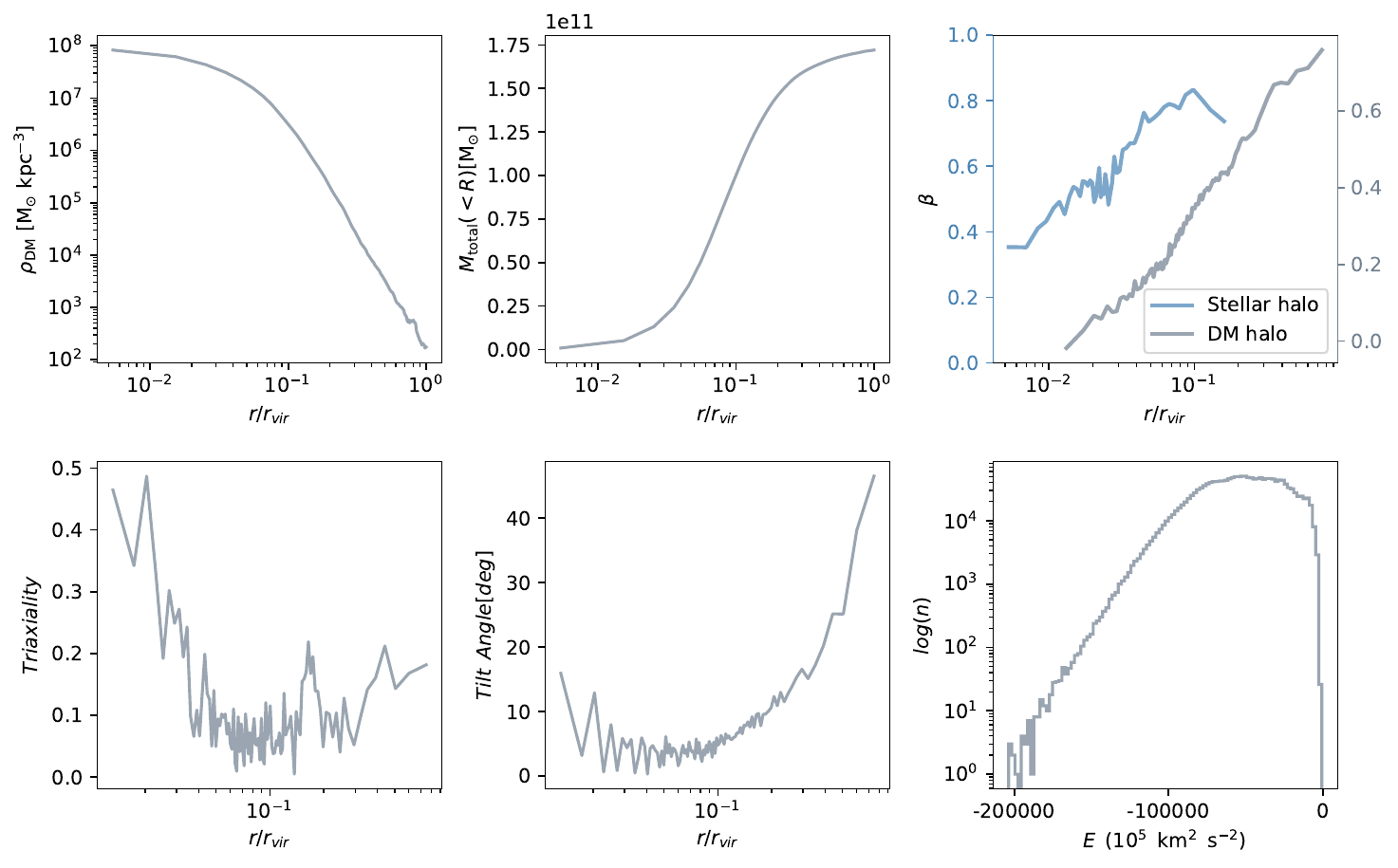}
      \caption{$Top\enspace left\enspace panel$ shows the DM density profile along with the radius. $Top\enspace middle\enspace panel$ shows mass enclosed within a given radius of DM halo. It can be observed that the mass of the dark matter halo in our model is concentrated within $\mathrm{0.2\,r_{vir}}$. $Top\enspace right\enspace panel$ is the anisotropy parameter $\beta$ of the stellar halo (blue solid line) and the DM halo (gray solid line) with Galactocentric distance in a spherical coordinate system. $Bottom\enspace left\enspace panel$ shows the triaxiality of the DM halo, which is measured with DM particles for each bins, as the radius is divided into 100 bins with the same number of particles from 0 to $\mathrm{r_{vir}}$, the dense polylines around $\sim \mathrm{0.1\,r_{vir}}$ exhibits a highly oblate shape. $Bottom\enspace middle\enspace panel$ is the tilt angle between DM halo's short axis with the disk plane in each radius bins, and it shows an increasing trend from the center outward.  $Bottom\enspace right\enspace panel$ shows hist profile of DM halo energy distribution within $\mathrm{r_{vir}}$.
              }
         \label{F2}
   \end{figure*}
    
   Due to the high eccentricity, the merger progresses rapidly, ta  king approximately 1 Gyr between the first and final pericenter passages and completing within the initial 3 Gyrs. In Figure~\ref{F1}, we present the rotation curve at 11 Gyr for the $15^{\circ}$ inclination case as an example, as differences between various inclination angle settings are negligible. Observational data from \citet{Jiao} and \citet{Ou} are included for comparison, demonstrating good agreement with our simulation. The rotation curve exhibits no flattening and shows a noticeable decline beyond $\mathrm{\sim 15\,kpc}$. Since the outer region of the RC mainly depends on the DM profile, additional details of the DM halo are provided in Figure~\ref{F2}. 

   The virial radius ($R_{\mathrm{vir}}$) of this MW model is about 131.05 kpc, defined as the radius within which the average dark matter density equals 200 times the critical density of the universe ($\rho_{\mathrm{cr}}$). This value is slightly higher than the 121.03 and 123.80 kpc reported by \citet{Jiao} and \citet{Ou}, respectively. The top-middle panel of Figure~\ref{F2} shows the enclosed mass within the virial radius ($R_{\mathrm{vir}}$) of the DM halo. Including the baryonic mass, the total mass is approximately $2.4\times10^{11}\,M_{\odot}$, comparable to the range of $2.44-2.53\times10^{11}\,M_{\odot}$ and $3.8-5.4\times10^{11}\,M_{\odot}$ reported by \citet{Jiao} and \citet{Ou}, respectively. Here, we measured the anisotropy parameter ($\beta$) for both stellar and DM halos, illustrating its radial variation using different colors. The parameter $\beta$ is defined as:
   \begin{equation}
    \label{f2}
      \beta = 1-\frac{\sigma_{\theta}^{2}+\sigma_{\phi}^{2} }{2\sigma_{r}^{2}}, \\
      \end{equation}
   Here, $\sigma_{i}$ represents the velocity dispersions in spherical coordinates. The stellar halo exhibits an increasing anisotropy trend at smaller radii, followed by a subsequent decline, similar to the observed features of GSE debris. In contrast, the anisotropy of the DM halo steadily increases up to the virial radius, indicating the orbits of DM particles become increasingly radial. We also determine the shapes and orientations of the principal axes of the halo using a standard iterative procedure based on the shape tensor method \citep{Emami}. In summary, particles are selected within the ellipsoidal volume defined by $\mathrm{r_{min}\le r_{ell}<r_{max}}$, where the ellipsoidal radius $\mathrm{r_{ell}}$ is defined as:
   \begin{equation}
   \label{f3}
        \left|r_{\mathrm{ell}}\right|^{2}=\left(\frac{r_{\mathrm{body}, 1}}{a}\right)^{2}+\left(\frac{r_{\mathrm{body}, 2}}{b}\right)^{2}+\left(\frac{r_{\mathrm{body}, 3}}{c}\right)^{2}, \\
      \end{equation}
      
    and
   \begin{equation}
   \label{f4}
        r_{\mathrm{body}} = E^{T}(r-r_{\mathrm{0}}). \\
      \end{equation}
      
    Here, $r_{\mathrm{body}}$ describes particle positions within the halo “body” frame (i.e., its position with respect to the halo principal axes). The halo center $r_{\mathrm{0}}$ is taken to be the position of the most bound halo particle (i.e., where the potential is lowest). The variables $a$, $b$, and $c$ are the halo shape parameters, and E gives the directions of the principal axes. We initialize these values with $a = b = c = 1$ and $E = I$, and update the values by diagonalizing the shape tensor whose elements are
   \begin{equation}
   \label{f5}
        S_{i,j} = \frac{1}{M}\sum_{k=1}^{N}m_{k}(r_{k})^{i}(r_{k})^{j}. \\
      \end{equation}
      
    The axis lengths $a$, $b$, and $c$ are proportional to the square root of the eigenvalues of the shape tensor, and they are normalized such that the ellipsoidal shell maintains a constant volume during iteration. The eigenvectors define the orientation of the halo principal axes $E$, and are updated each iteration. At every step, the halo shape is computed as the ratio of the minor-to-major axes ($q=c/a$), and the ratio of the intermediate-to-major axes ($p=b/a$). The iteration process is terminated when the residual of both tow ratios converges to a level below:
   \begin{equation}
   \label{f6}
        MAX(((q-q_{old})/q)^2,((p-p_{old})/p)^2)\le 10^{-3}. \\
      \end{equation}    

    Where MAX refers to the maximum between the two quantities. Then we can calculate the triaxiality with $T=(1-p^{2})/(1-q^{2})$. $T>0.6$ denotes a prolate halo, and $T<0.3$ indicates an oblate halo. By measuring the orientation of the axes (with the disk plane's normal vector aligned parallel to the Z-axis), we are able to determine the tilt angle. Our measurement indicate that the major and intermediate axes are degenerate, a scenario also observed in \citet{Kazantzidis} and \citet{Shao}. This degeneracy renders the measurement unstable if these axes are used. However, the minor axis is nondegenerate and can be reliably used to measure the tilt angle of the DM halo with respect to the disk plane. 
    
    In the bottom panel of Figure~\ref{F2}, we analyze the DM halo by this method, dividing the radius into 100 bins, each containing an equal number (17191) of DM particles. The dense polylines are concentrated within $\mathrm{0.2\,r_{vir}}$, indicating that the inner DM halo is more prolate, which corresponds to the formation of a dark matter bar. Meanwhile, in the densest region, the halo exhibits a highly oblate shape. The triaxiality varies only slightly with increasing radius. The tilt angle of each bin shows an increasing trend from the center outward. However the angles are highly consistent within $\mathrm{0.2\,r_{vir}}$, indicating that the dark matter halo is compact. The non-zero triaxiality and the radial variation of the tilt angle confirm that the halo is triaxial and misaligned with the disk.

    We also apply the same method to measure the time evolution of triaxiality and tilt angle under different inclination settings. We selected DM particles within $\mathrm{20\,kpc}$, a region that contains the majority of the galactic mass. In the top panel of Figure~\ref{F3}, we plot triaxiality over simulation time from 3 to 11 Gyr. The inclination angles are divided into two groups: high ($i>45^{\circ}$) and low ($i\le 45^{\circ}$). In the early stages, the high-inclination group exhibits larger triaxiality, even approaching prolate shapes. However, in the later stages, DM halos in all inclination settings tend to become oblate. Additionally, simulations with $i=15^{\circ},30^{\circ},45^{\circ}$ display quasi-periodic evolution in halo shape. Figure~\ref{F3} bottom panel shows the evolution of the tilt angle relative to the disk plane, derived by tracking the short axis of the DM halo. In the high-inclination group ($i=75^{\circ},60^{\circ},55^{\circ}$ also in the $45^{\circ}$), the tilt angle varies significantly over time, whereas in the low-inclination group ($30^{\circ},15^{\circ}$), the evolution is more stable. Nonetheless, even subtle changes in the tilt angle can manifest as gravitational disturbances in the galactic disk, which will be discussed later.
    
   \begin{figure*}
   \label{F3}
   \centering
      \includegraphics[scale = 0.55]{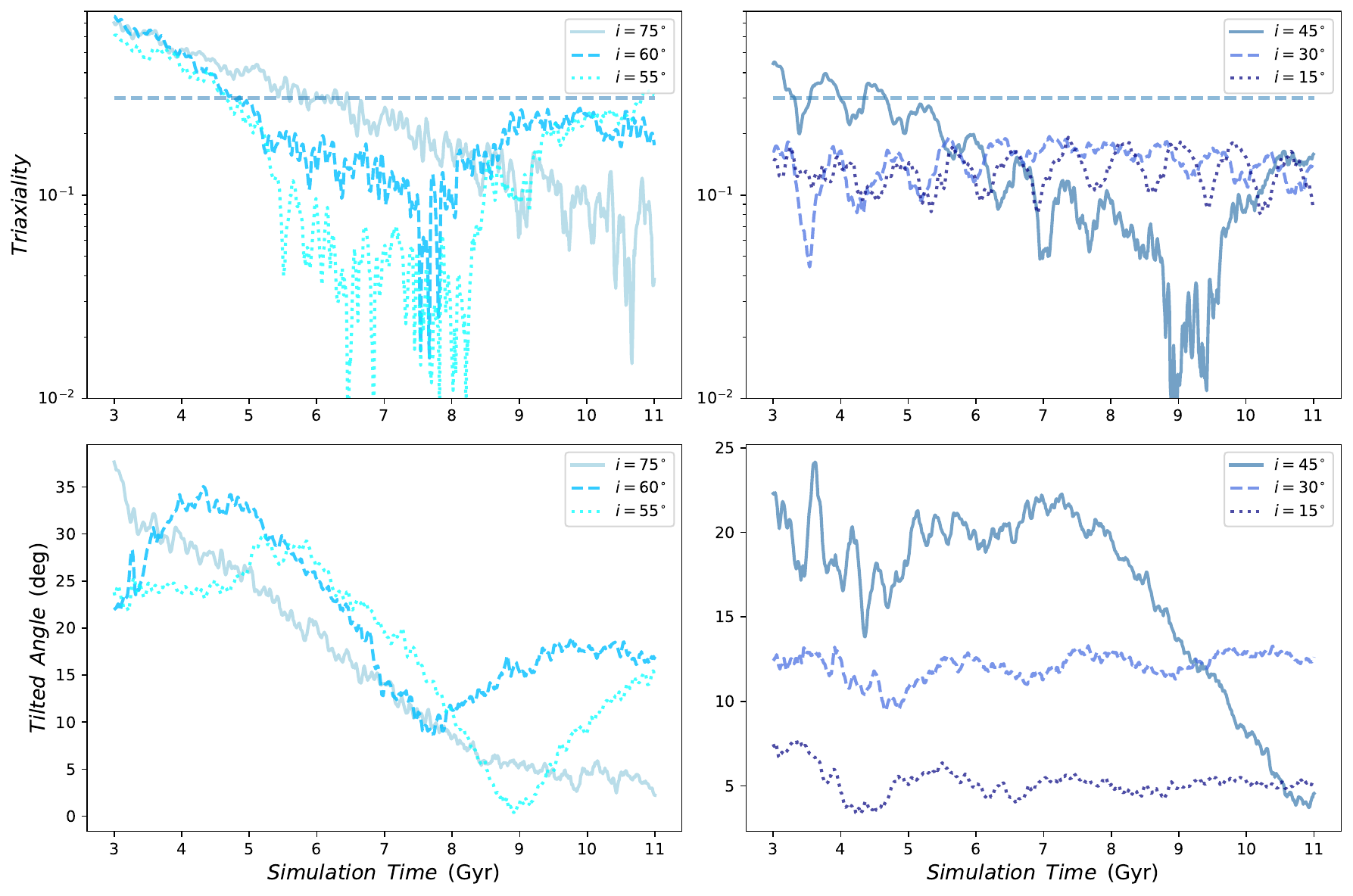}
      \caption{Time evolution of the DM halo's shape and orientation for different inclination groups (indicated by different colors and lines in each subgraph), calculated using DM particles within the central $20$ kpc . $Top\enspace panels$ show the evolution of the DM halo's triaxiality parameter, defined as $T=(1-p^{2})/(1-q^{2})$, from 3 to 11 Gyr. The horizontal dashed line marks $T=0.3$; values below this threshold indicate an oblate shape, while values $T\geq0.6$ indicate a  prolate shape.  $Bottom\enspace  panels$ indicate the evolution of the DM halo's tilted angle with respect to the disk panel, derived by tracking the short axis of the DM halo. The halo evolution shows distinct behaviors roughly based on inclination: high-inclination models ($i>45^{\circ}$)  , and low-inclination models ($i\leq45^{\circ}$), they both retain tilted with respect to the disk all the time.
              }
         \label{F3}
   \end{figure*}

   \section{Galactic vertical Waves} \label{sec: Gv w}
   \subsection{Evolution of disk warp}
    The geometric shape of the warp can be approximated by a power-law warp or $\mathrm{m=1}$ mode:
    
      \begin{equation}
      \label{f7}
      Z_{w}(R\ge R_{w}) =a(R-R_{w})^{b} sin(\phi -\phi_{w});\\
      \end{equation}
      \begin{equation}
      \label{f8}
      Z_{w}(R< R_{w}) =0.\\
      \end{equation}
   
   Here, $R$ and $Z_{w}$ are Galactocentric cylindrical coordinates, where $a$ represents the amplitude of the warp, $b$ is the power-law index, $\phi_{w}$ is the polar angle of the warp's line of nodes (LON hereafter) indicates the orientation of the warp and $R_{w}$ is the onset radius of the warp. This model was employed by \citet{Chen} to accurately fit the Cepheid tracers, prompting us to select stars within the young stars' age range in the simulation for comparison with the Cepheid data. As \citet{Chen} note, the parameters of $a$, $b$ and $R_{w}$ show a clear correlation, whereas $\phi_{w}$ is a more independent parameter in the geometric warp model. Consequently, while different studies might yield varying $a$, $b$ and $R_{w}$ values, $\phi_{w}$ tends to be more consistent across measurements of the Galactic warp. 

   \begin{figure*}
   \label{F5}
   \centering
      \includegraphics[scale = 0.55]{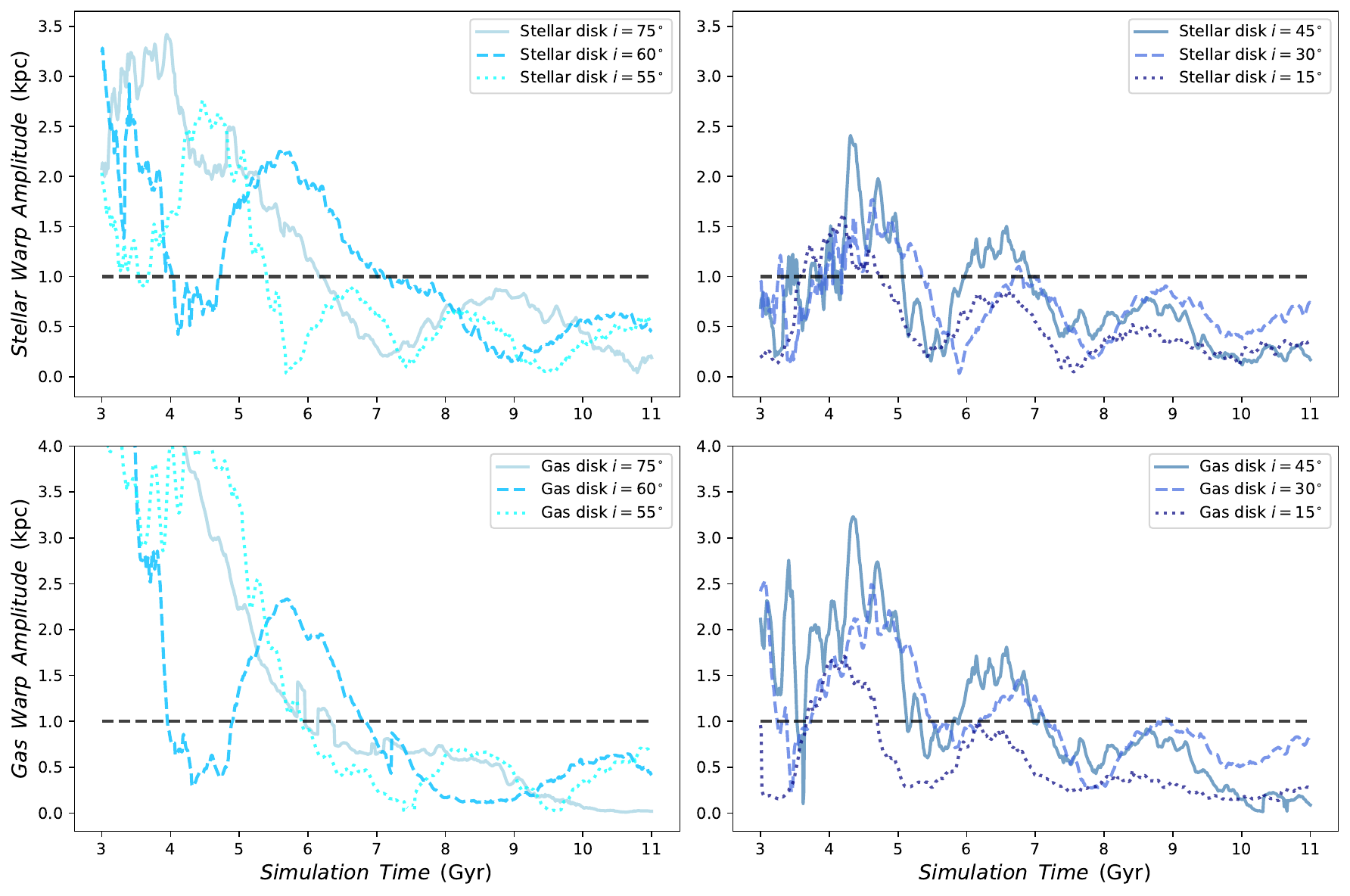}
      \caption{$Top\enspace panels$ show the evolution of stellar warp and the $Bottom\enspace panels$ indicate the gas disk warp amplitude, both measured at R = 16 kpc, from 3 Gyr to 11 Gyr, and the stellar disk amplitude is fitting with stars younger than 1.5 Gyr. The curves represent different inclination models as defined in Figure~\ref{F3}. The dash lines indicate the amplitude of MW in observation. High-inclination models show larger initial amplitudes, and the warp displays a regeneration phenomenon: the amplitude often decays to a near-flat state before rising again to a new peak.              }
         \label{F5}
   \end{figure*}
   
   Figure~\ref{F5} illustrates the time evolution of the warp amplitude in the stellar and gas disks. The amplitude is calculated by fitting Eq.~\ref{f7} to the warp every 0.01 Gyr starting from $t = 3$ Gyr—after the merger is completed—and identifying the maximum amplitude at $R = 16$ kpc, which is near the simulated disk edge. The warp is measured using young disk particles selected based on circularity $\epsilon = L_{Z}/L_{Z,\text{max}(E)} > 0.65$ and age less than 1.5 Gyr in the simulation. This selection reflects that the warp is most prominent among the youngest stars and spans multiple age populations \citep{Chen, Cheng, Huang}. The dashed lines in both panels represent the observed amplitudes of the Milky Way’s stellar and gas disk warps \citep{Chen, Levine}. We specifically focus on this young population for three reasons. First, older stellar populations (e.g., $\leq$ 4 Gyr) exhibit significant vertical dispersion, which introduces noise into the warp signal \citep{Deng, Han23}. Second, observational studies indicate that warp kinematics derived from young tracers, such as Cepheids ($\sim100$ Myr), show consistent constraints on the precession rate and LoN orientation, whereas results from older tracers remain inconclusive \citep[see Figure 8 in][]{CC}. Finally, we tested the robustness of our selection by comparing it with a broader sample including stars up to 4 Gyr. We found that in the early stage, when the proportion of old stars is high, the measurements will deviate from those of young stars. However, as the proportion of old stars decreases, the two measurements will gradually become consistent. Regarding the measurement of precession, we found that there is not much difference between the two sampling methods. Therefore, the specific choice of sampling does not affect our final conclusions. Overall, we maintain that choosing young stars ensures the robustness of our results while being more realistic with observations.

   Since the Galactic stellar and gas disks exhibit similar warp amplitudes, the stellar and gas disks within the same inclination group also show comparable evolutionary trends in Figure~\ref{F5}. This suggests they are dominated by the same evolutionary mechanism or influenced by the same gravitational potential. In high-inclination groups, the warp amplitude is larger during the early stages, possibly due to the higher vertical angular momentum of the merger orbit. Most notably, a regeneration mechanism appears in the warp evolution. At times, the disk warp amplitude declines to a near-flat state, only to regenerate and rise again to a new peak. Since the simulation involves only a single initial merger event, this persistent, quasi-periodic regeneration provides strong evidence that the warp is driven by long-lived internal processes.

\subsection{Kinematic warp model} \label{sec:Kinematic warp model}

   In the previous subsection, we examined the evolution of the Galactic warp, emphasizing that its rapid changes in amplitude reveal the inherently dynamical nature of warps. Understanding this dynamism is essential for tracing the formation history of galaxies and determining the mass distribution of their halos. The changing geometry of the warp can be characterized by the variation in the direction of its lines of nodes (LONs), which precess at a rate $\omega$. This relationship is given by $\phi_{w}(t) = \phi_{w,0}+\omega t$ where $\phi_{w,0}$ denotes the current position of the LONs. A time-dependent warp model based on this framework is presented by \citet{Poggio}:
      \begin{equation}
      \label{f9}
      \overline{V_{Z}}(R,\phi,t=0)=(\frac{\overline{V_{\phi}}}{R} -\omega(R))h_{w}(R)cos(\phi-\phi_{w})+\frac{\partial h_{w}}{\partial t}sin(\phi-\phi_{w}).\\
      \end{equation}
      
   Here $h_{w}(R)=a(R-R_{w})^{b}$, as described in Eq.~\ref{f1}, $\overline{V_{Z}}$ represents the mean vertical velocity and $\overline{V_{\phi}}$ represents the mean azimuthal velocity. To simplify the calculations, we neglect the time derivative term $\frac{\partial h_{w}}{\partial t}$ and minimize the length of the $R$ bin and assume that $\omega$ remains constant with respect to $R$. Although the warp is evolving, our calculations show that this approximation has little effect on the precession results. Under these assumptions, the model can be expressed in the same form as Equation 7 in \citet{Poggio}:
      \begin{equation}
      \label{f10}
      \overline{V_{Z}}(R,\phi)=(\frac{\overline{V_{\phi}}}{R} -\omega)h_{w}(R)cos(\phi-\phi_{w}).\\
      \end{equation}
      
   Using this kinematic model, once the geometric shape is determined by fitting Eq.\ref{f7}, we can substitute the parameters $h_{w}$ and $\phi$ to measure the precession rate $\omega$ based on the simulation’s kinematic data. For this analysis, we select young stars ($\mathrm{age<1.5\,Gyr}$) within the radial range $R=7-17\,\text{kpc}$. The region is then divided into 20 radial bins to track and plot the temporal evolution of $\omega$.

   \begin{figure*}
   \label{F6}
   \centering
      \includegraphics[scale = 0.55]{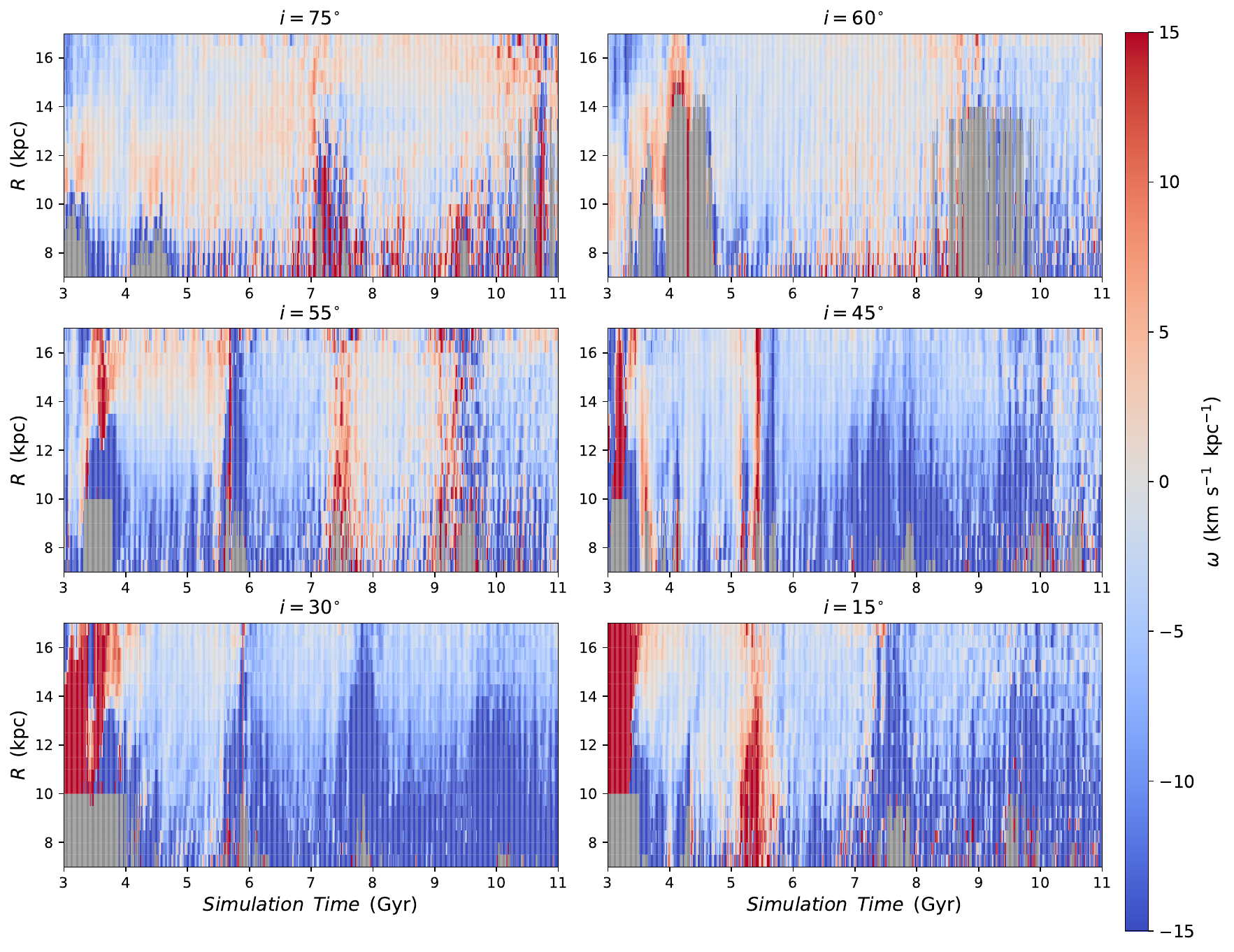}
      \caption{Evolution of stellar disk warp precession $\omega$ of the higher and lower inclination groups from $\mathrm{3\,Gyr}$ to $\mathrm{11\,Gyr}$ as measured with young stars within the radial range of $\mathrm{R = 7-17\,kpc}$, which are divided into 20 bins at intervals of $\mathrm{0.5\,kpc}$. The color means the direction of precession, while red means prograde ($\omega\,\textgreater\,0$), blue means retrograde ($\omega\,\textless\,0$) and the white color indicates the non-precession state. Some bins are colored in gray to represent the non-warp region where $R < R_{w}$. Here, $R_{w}$ is the onset radius of the warp. The kinematic behavior differs by inclination: high-inclination models frequently exhibit long-term prograde precession, while low-inclination models are dominated by retrograde precession with only transient high prograde features.              
              }
         \label{F6}
   \end{figure*}

   Figure~\ref{F6} shows time evolution of stellar disk precession across different inclination groups. Long-term prograde precession is more frequently observed in the high-inclination group, although transient high prograde precession is also occasionally observed in the low-inclination group, it appears as a temporary feature and is generally dominated by retrograde precession for the majority of the simulation time. The disk warp can be described as an $\mathrm{m=1}$ bending wave propagating in a kinematically cool disk. Such a wave can be viewed as the superposition of ‘fast’ ($+$) and ‘slow’ ($-$) waves circulating with frequencies $\omega = m\Omega(R)\pm \nu(R)$, where $\Omega(R)$ is the angular rotation curve, $\nu(R)$ is the vertical frequency, and m-fold rotational symmetry is assumed. The inclusion of the disk’s selfgravity raises $\nu(R)$, making bending waves more stable, i.e. stiffer (contrary to density waves) – see \citet{BT}. For a given rotation curve $\omega(R)$, bending waves can only propagate in regions that satisfy the condition
      \begin{equation}
      \label{f11}
      \left [  \Omega_{P}- \Omega(R)\right ]^{2}\ge  \nu_{h}^{2}.\\
      \end{equation}
      
   where $\Omega_{p}=\omega/m$ is the pattern speed and $\nu_{h}$ is the frequency of vertical oscillation contributed by the halo potential. This defines, for m = 1, a ‘forbidden’ region,  $\Omega - \nu_{h}<\Omega_{p}<\Omega + \nu_{h}$, where bending waves cannot propagate. Generally, $\nu>\Omega$, so the ‘fast’ wave is prograde, with a frequency $\omega$ depending strongly on $R$ for most radii, so differential rotation winds it up rapidly and it decays. The ‘slow’ wave, on the other hand, is retrograde and circulates with frequency only weakly depending on $R$ for most radii. This wave is thus expected to wind up slowly and be long-lived. In other words, prograde waves generally disappear quickly, except for persistent disturbances.

   In addition, according to Eq.~\ref{f10}, a positive $\omega$ (prograde precession) leads to a reduction in the vertical velocity of the disk. Conversely, retrograde precession increases the vertical velocity. Therefore, when the disk is in a high prograde but transient precession phase, the warp amplitude often reaches a minimum in a short time of its evolution, and a low prograde but long-lasting precession, the warp amplitude is in the process of continuous attenuation For example, these behaviors can be seen during the periods of $\mathrm{3-3.5\,Gyr}$ and $\mathrm{5-5.5\,Gyr}$ in the $15^{\circ}$ inclination, and during the period of $\mathrm{3-11\,Gyr}$ of the $75^{\circ}$ inclination in Figure \ref{F6} respectively.
   
  \subsection{Perturbation from the DM halo}\label{sec: Perturbation from the DM halo}
   As discussed previously, the disk appears to be influenced by a gravitational disturbance. In our simulation, however, the only source of disturbance is the DM halo. To quantify the impact of the dark matter halo on the vertical structure of the disk, we calculated the vertical gravitational acceleration $a_{z}$ exerted on both the $Z=0$ plane and the actual warped disk plane (which typically deviates from $Z=0$ near the edges) at each simulation snapshot with the DM particles within the $\mathrm{R\leq20\,kpc}$  range. This is computationally faster than calculating accelerations on all disk particles. This mesh has $100\times100$ bins in the X and Y directions and extends to $\mathrm{ 17\,kpc}$ ($\mathrm{X,Y\subseteq \left [  -17,17\right ]\, kpc}$). We used the following expression
      \begin{equation}
      \label{f12}
      a_{z, \text { bin }}=\sum_{i}^{N_{\text {part }}} \frac{-G m_{i}}{r_{i, \text { bin }}^{2}} \frac{r_{z ; i, \text { bin }}}{r_{i, \text { bin }}},\\
      \end{equation}
   where the vertical acceleration at each bin ($a_{z,\text{ bin }}$) is the sum of all individual vertical accelerations from all the mass elements $N_{part}$ (dark matter particles). $m_{i}$ is the mass of each mass element, and $r_{i,\text{ bin }}$ is the total distance between the bin and the mass element. The relation $r_{z;i,\text{ bin }}/r_{i,\text{ bin }}$ takes the vertical component of the acceleration. In the same range, we also calculate the average coordinates and velocity of the thin disk in the vertical direction.

   Figure~\ref{F8} illustrates the evolution of the warp amplitude, transitioning from a high state to a low state and then back to a high amplitude. In the top panel ($\mathrm{t=4.19\,Gyr}$), the warp reaches a crest. Vertical displacement, velocity, and acceleration are prominent at the disk edge, while the acceleration on the $Z=0$ plane does not exhibit the strong bipolar features seen in the disk plane. In the middle panel ($\mathrm{t=5.43\,Gyr}$), while the disk appears nearly flat ($\overline{Z} \approx 0$) and the acceleration on the disk particles is weak, the acceleration on the $Z=0$ plane remains obvious dipole structure, which indicates that the external gravitational potential which is driven by the DM halo continues to exert torque on the disk even when the warp is visually absent, forcing the disk to warp again. Then in the bottom panel $\mathrm{t=6.17\,Gyr}$, the warp reaches another crest, with a smaller amplitude compared to that at $\mathrm{t=4.19\,Gyr}$. The phases of displacement and acceleration on the disk are found to be consistent. The velocity phase leads by approximately $90^\circ$, meaning that velocity peaks where displacement is zero—consistent with the wave-like nature of the warp. 
   
   To visualize the temporal evolution of bending parameters, we extract the vertical displacement, velocity, and acceleration values from both planes using an X–Y grid. These values are then mapped into a polar coordinate system and analyzed via Fourier decomposition, with the plane divided into radial sections of $\mathrm{0.34 \,kpc}$ in width. Each radial ring contains N bins from the original X–Y grid. Each bin is located at an azimuthal position $\phi_{i}$ and has a bending amplitude (weight in the Fourier decomposition) of $W_{i}$ ($W_{i}=Z_{i},V_{Zi},a_{zi}$). We then perform a Fourier decomposition using the following expression:
      \begin{equation}
      \label{f13}
      \mathrm{a_{m} = 2\sum_{i}^{N}W_{i}cos(m\phi_{i})},\\
      \end{equation}  

      \begin{equation}
      \label{f14}
      \mathrm{b_{m} = 2\sum_{i}^{N}W_{i}sin(m\phi_{i})},\\
      \end{equation} 

     \begin{equation}
      \label{f15}        \mathrm{A_{m}=\sqrt{a_{m}^2+b_{m}^2}/N}.\\
      \end{equation}

   \begin{figure*}
   \label{F8}
   \centering
      \includegraphics[scale = 0.55]{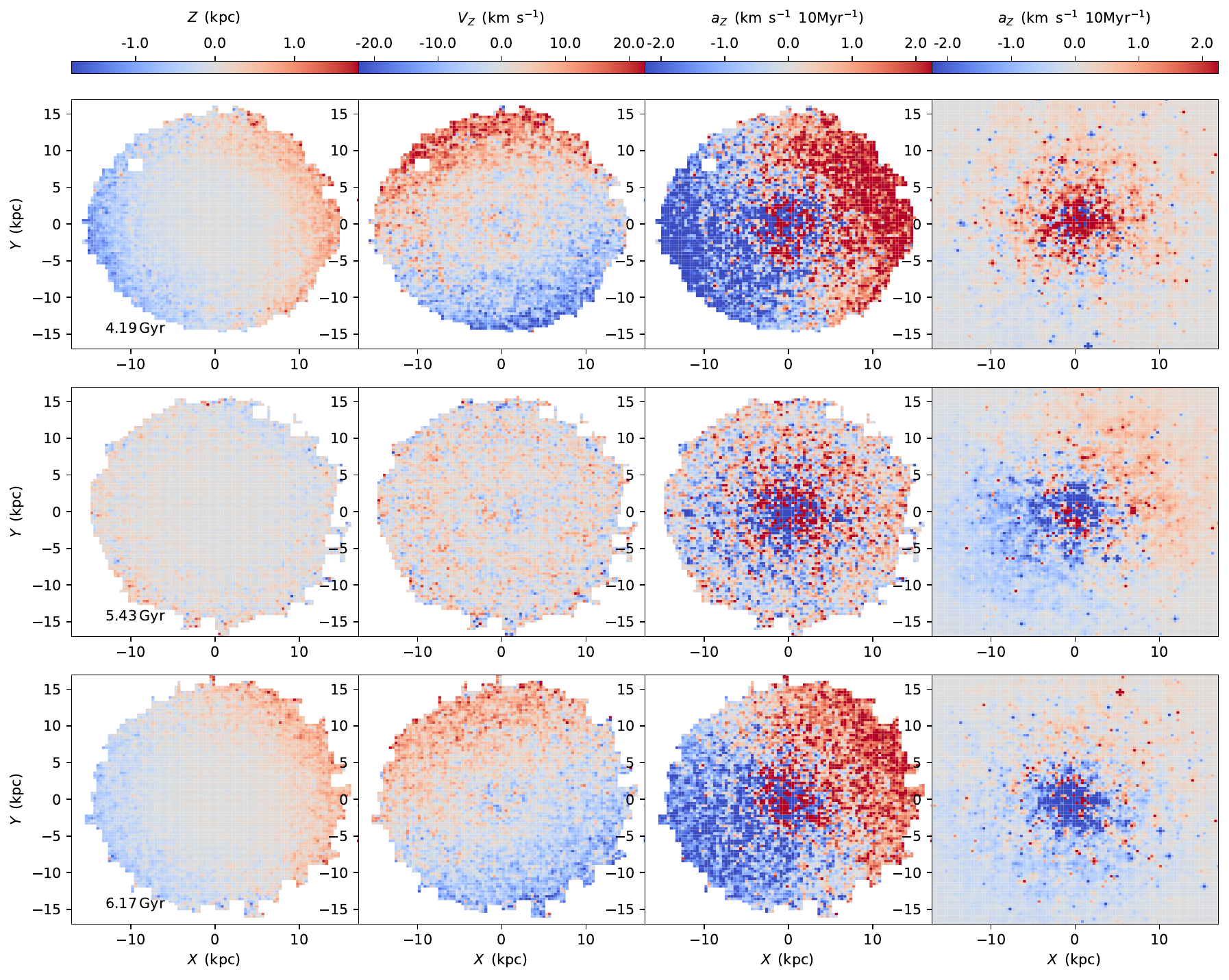}
      \caption{ Two-dimensional maps of the stellar disk's vertical structure and kinematics in the $X-Y$ plane (spanning $\pm 17$ kpc) for the $15^{\circ}$ inclination model. The snapshots capture a full cycle of warp regeneration at three distinct epochs: $Top\enspace panels$: $t=4.19$ Gyr (warp crest amplitude);  $Middle\enspace panels$: $t=5.43$ Gyr (warp trough amplitude);  $Bottom\enspace panels$: $t=6.17$ Gyr (regenerated crest). $Columns$ (Left to Right): (1) average vertical displacement $\overline{Z}$; (2) average vertical velocity $\overline{V_{Z}}$; (3) vertical acceleration experienced on disk plane $\overline{a_{Z}}$ (disk); (4) vertical acceleration measured on the $Z=0$ plane $\overline{a_{Z}}$ ($Z=0$). The color depth represents the magnitude, and the color hue indicates the direction (positive/negative), with acceleration units in km s$^{-1}$ 10Myr$^{-1}$ here 10Myr$^{-1}$ is the time step of our simulation. This
      sequence visualizes the warp's regeneration mechanism.              }
         \label{F8}
   \end{figure*}
  We estimated the amplitudes $A_{m}$ of each parameter at every radius and time step. In the subsequent analysis, we focus on the $m=1$ mode. Other modes were also examined, but they were found to be subdominant. We also examine the phase angle at each radius, calculated using the following expression:
     \begin{equation}
      \label{f16}        \mathrm{\phi_{m}=arctan(b_{m}/a_{m})}.\\
      \end{equation}

   \begin{figure*}
   \label{F9}
   \centering
      \includegraphics[scale = 0.4]{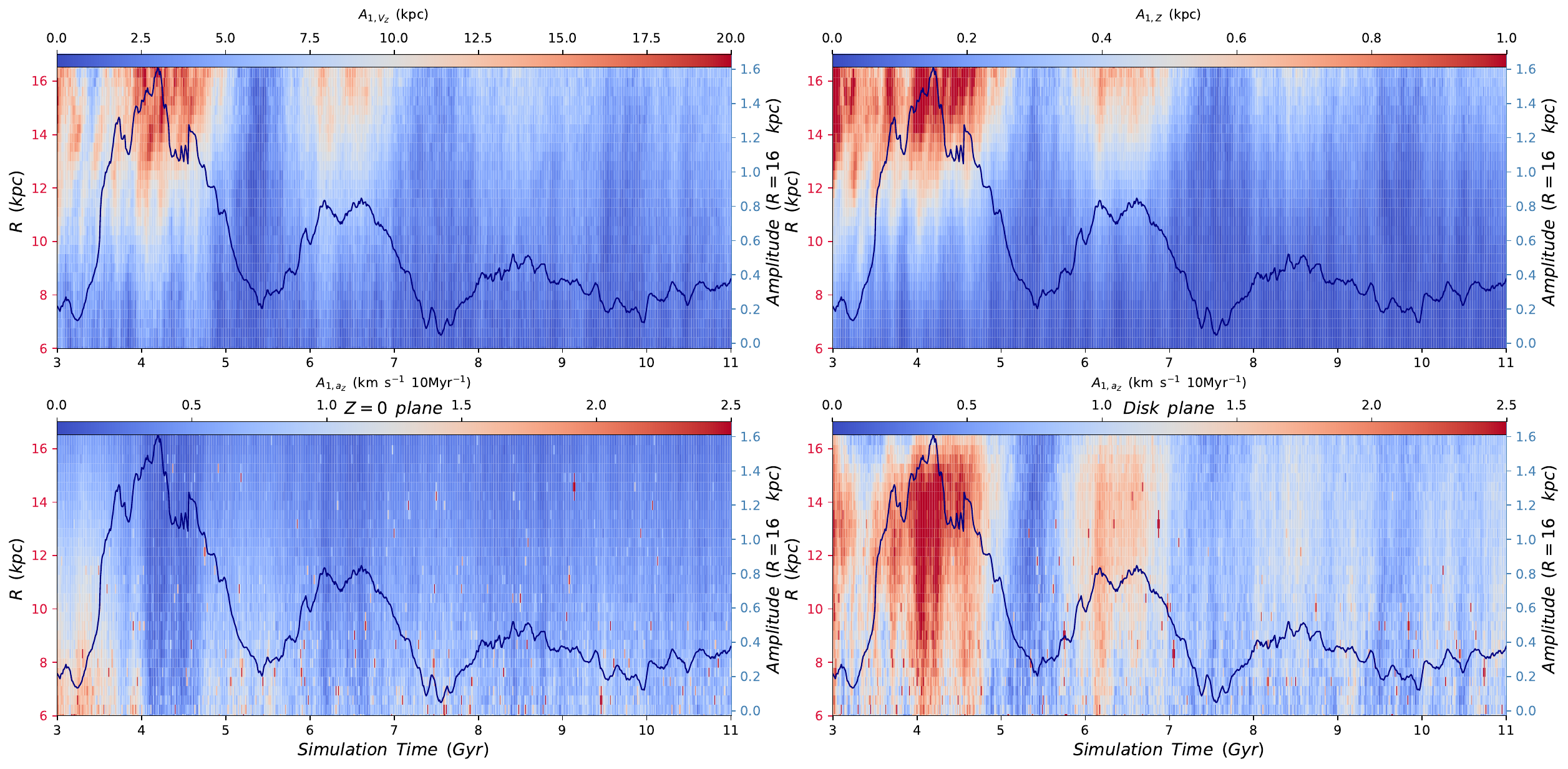}
      \caption{Temporal and radial evolution of the $m=1$ Fourier bending mode amplitudes, quantifying the strength of the warp. The color maps display the Fourier amplitude $A_1$ as a function of time (horizontal axis) and radius (left vertical axis). The overlaid curve in deep blue (corresponding to the right vertical axis) traces the global warp inclination amplitude for comparison. $Top\enspace left\enspace panel$ is the vertical velocity amplitude $A_{1, V_Z}$; $Top\enspace right\enspace panel$ is the  vertical displacement amplitude $A_{1, Z}$. $Bottom\enspace left/right\enspace panel$ is the vertical acceleration amplitude $A_{1, a_Z}$ measured on the $Z=0$/disk plane. The maps reveal the global, coherent nature of the bending waves.           }
         \label{F9}
   \end{figure*}

   \begin{figure*}
   \label{F10}
   \centering
      \includegraphics[scale = 0.4]{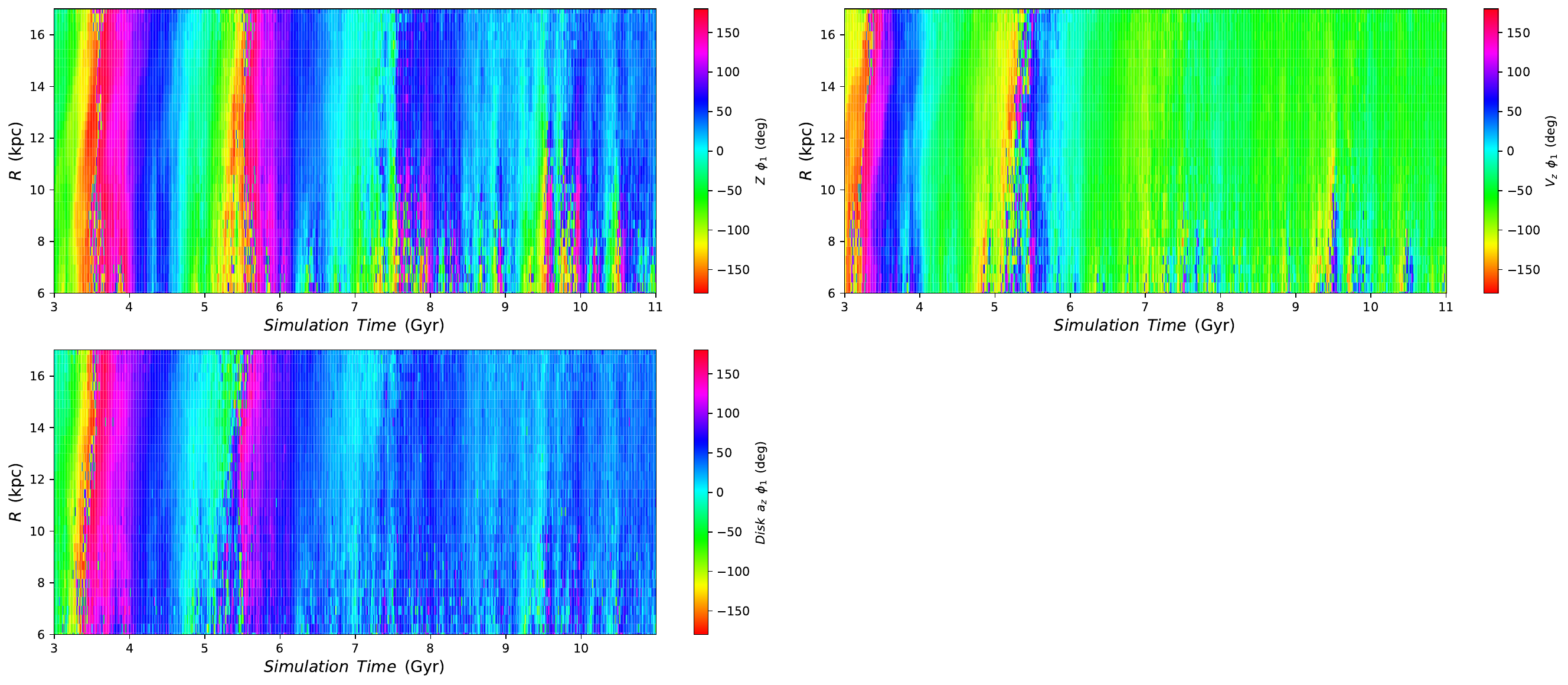}
  \caption{Temporal evolution of the phase angle $\phi_1$ for the $m=1$ bending mode, mapped as a function of time (horizontal axis) and radius (vertical axis). The color scale represents the phase angle in degrees, ranging from $-180^\circ$ to $180^\circ$. $Top\enspace left\enspace panel$ shows phase of vertical displacement $\phi_{1, Z}$; $Top\enspace right\enspace panel$ is the phase of vertical velocity $\phi_{1, V_z}$; $Bottom\enspace panel$ is the phase of vertical acceleration $\phi_{1, a_Z}$ on the disk plane. These phase maps provide a  dynamical confirmation of the warp as a propagating bending wave. Consistent with wave theory, the displacement phase and acceleration phase (two left panels) are largely coherent in phase, the velocity phase  exhibits a systematic phase offset of approximately $90^\circ$ relative to the acceleration.              
              }
         \label{F10}
   \end{figure*}
   
   Figure~\ref{F9} shows the amplitude of the bending mode $m=1$ as a function of radius and time. For comparison, the evolution of the warp amplitude from Figure~\ref{F5} is also overplotted on each panel. During the early strong perturbation phase ($3\text{--}5$ Gyr), velocity amplitudes reach $\sim 20$ km s$^{-1}$ in the outer radii. Crucially, the peaks in the vertical acceleration amplitude on the disk plane (around $t=4$ Gyr) align temporally with the maxima in the warp amplitude (overlaid curve) and the kinematic response. This synchronization confirms that the time-varying gravitational field of the dark matter halo is the driving force affecting the disk warp. Subsequent peaks show declining intensities (lighter colors), indicating the gradual damping of the bending waves, yet the correlation between the external halo torque and the disk's response remains persistent.

   According to wave theory, the displacement of a particle should be in phase with the force it experiences, while its velocity should lead or lag by $\pi/2$. Such relationship is illustrated in Figure~\ref{F10}. For most of the simulation time, the displacement and acceleration phases are consistent, and the trend is especially clear before 7.5 Gyr. After, the coherence weakens, possibly due to noise or numerical artifacts. However, several distinct phase-coherent features can still be identified, particularly during $\mathrm{7.5-8.5\,Gyr}$ and  $\mathrm{9.5-10.5\,Gyr}$. This inconsistency may stem from limited spatial resolution. The gravitational field was calculated on a $100 \times 100$ grid, whereas displacement and velocity were computed using all thin disk particles within the same spatial range. The mismatch in spatial resolution and the dispersion of disk particles likely contribute to phase inconsistencies in the later stages, where the warp amplitude also decreases, as shown in Figure~\ref{F5}. The velocity phase of Figure~\ref{F10} shows a phase offset of approximately $90^\circ$ relative to the acceleration. For example, regions with phase $\sim -50^\circ$ (green) in the velocity map correspond to phase $\sim 50^\circ$ (blue) in the acceleration map. Additionally, the temporal evolution of the two phases shows similar trends. For instance, around $\mathrm{6.9\,Gyr}$, both acceleration and velocity phases are decreasing, from green ($\sim-50^{\circ}$) to yellow ($\sim-100^{\circ}$) for the velocity phase and acceleration phase changes from blue ($\sim50^{\circ}$) to light blue ($\sim0^{\circ}$), while near $\mathrm{\sim8\,Gyr}$, both begin to increase simultaneously, from green to light blue for the velocity and from blue to deep blue ($\sim75^{\circ}$).

   \begin{figure*}
   \label{F11}
   \centering
      \includegraphics[scale = 0.4]{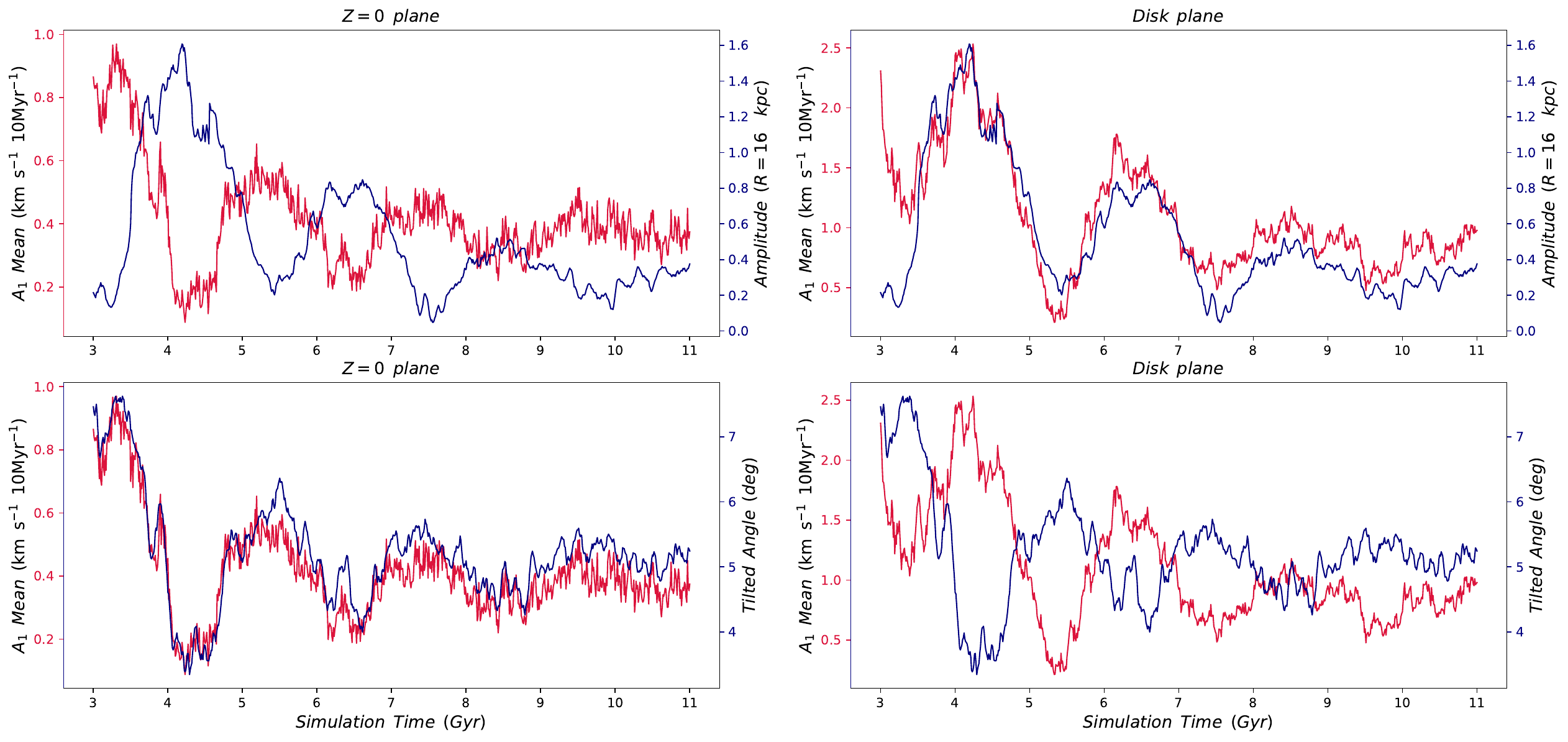}
  \caption{Correlations between the DM halo's gravitational field, its geometric orientation, and the disk warp evolution. The red curves represent the mean amplitude of the $m=1$ vertical acceleration component $A_{1, a_Z}$ (left vertical axis). averaged over the outer disk ($R=8\text{--}17$ kpc). $Top\enspace panels$ show the time evolution of $A_{1, a_Z}$ measured on the $Z=0$ plane ($left$) and the disk plane ($right$). The blue curves trace the warp amplitude (right vertical axis). $Bottom\enspace panels$ is the time evolution of the DM halo's tilt angle (blue curves, right vertical axis) compared against the same acceleration amplitudes ($A_{1, a_Z}$, red curves) from the top panels. The acceleration on the $Z=0$ plane is notably out of phase with the warp amplitude, and the driving acceleration is directly governed by the halo's geometry, the acceleration on the $Z=0$ plane correlates strongly with the halo's tilt angle. This temporal variation of the halo's orientation induces the time-varying gravitational torque that powers the warp's regeneration through this angular momentum exchange.
              }
         \label{F11}
   \end{figure*}

  In the top panel of Figure~\ref{F11}, we focus on the radial range $\mathrm{R\subseteq \left [  8,17\right ]}$, within which we calculate the mean amplitude of the  $m=1$ component of the vertical acceleration on different planes. The inner region is excluded because some mass elements lie very close to the plane, producing artificially high accelerations and increasing noise in the Fourier analysis. Moreover, bending waves are more dominant in the outer disk. For comparison, we also plot the warp amplitude, which shows clear consistency with the acceleration amplitude. However, the amplitude of acceleration on the $Z=0$  plane is out of phase with the warp amplitude: when one reaches its maximum, the other tends to be at a minimum. Since the acceleration is derived from the DM halo, this suggests that warp evolution depends on the evolution of the halo itself. When the gravitational field from the DM halo strengthens in a particular direction, disk particles respond by moving accordingly, leading to the formation of a warp. Conversely, if the gravitational field shifts back toward where $Z=0$,  the warp gradually diminishes.

  The characteristic feature of an $m=1$ warp mode is that one edge of the disk bends upward while the opposite edge bends downward. Therefore, the gravitational field provided by the DM halo must increase in opposite directions simultaneously to generate this effect. This behavior suggests a tilted ellipsoidal DM halo. If the halo is inclined relative to the disk plane, it naturally generates an asymmetric gravitational potential. If the DM halo undergoes rotation or other forms of motion that alter its tilt angle, the resulting changes in gravitational alignment will affect the strength of the disk disturbance and the warp amplitude. 
  
  In the bottom panel of Figure~\ref{F11}, we plot the acceleration amplitude on different planes as a function of the tilt angle derived in Figure~\ref{F3}. Although this angle was previously described as the tilt relative to the disk plane, it is difficult to define a consistent disk plane due to continuous warping at the edges. Therefore, we measure the tilt with respect to the $Z=0$ plane. The bottom left panel shows that the tilt angle correlates strongly with the acceleration amplitude on the $Z=0$ plane, indicating that the halo’s tilt angle relative to the $Z=0$ plane increases, the vertical acceleration on that plane also increases, and confirming that a larger halo tilt generates a stronger perturbative field. However, the bottom right panel of Figure~\ref{F11} shows an inverse relationship between acceleration amplitude on the disk plane and tilted angle: as the acceleration on the disk plane increases, the tilted angle decreases.

  The observed anti-correlation between the acceleration amplitude on the $Z=0$ plane and the warp amplitude can be understood through the conservation of total angular momentum. The galaxy-halo system acts as a set of coupled oscillators. Since the total angular momentum vector $\vec{L}_{\text{tot}} = \vec{L}_{\text{disk}} + \vec{L}_{\text{halo}}$ is conserved, any increase in the halo's misalignment (tilt angle $\theta_{\text{halo}}$) must be balanced by a decrease in the disk's misalignment on the edge: the warp ($\theta_{\text{disk}}$), weighted by their respective angular momenta ($L_{\text{halo}} \theta_{\text{halo}} + L_{\text{disk}} \theta_{\text{disk}} \approx \text{const}$). Since the vertical acceleration on the $Z=0$ plane, $a_{Z=0}$, serves as a direct proxy for the halo's tilt $\theta_{\text{halo}}$, its maximum naturally coincides with the minimum of the disk's warp amplitude $\theta_{\text{disk}}$. This `seesaw' mechanism confirms that the warp evolution is driven by the continuous exchange of angular momentum between the misaligned DM halo and the disk. Over time, the angular momenta of the halo and the disk tend to align, reducing the amplitude of the halo's tilt oscillation. Consequently, the amplitude of each subsequent regenerated warp also diminishes.
  
  These results all above suggest that the motion (e.g., rotation or precession) of the DM halo alters its tilt angle relative to the disk, leading to a misalignment between the gravitational field and the disk plane, which in turn induces a warp. Furthermore, temporal variations in this misalignment can lead to a time-varying warp with precessional characteristics.

  In \citet{Han23}, a fixed tilt angle between the DM halo and the disk is assumed. After approximately 5 Gyr of evolution, the simulation reproduces a MW-like disk warp. However, the resulting warp is long-lived and steady, lacking precession, and thus deviates significantly from realistic physical scenarios. As noted above, the tilt angle relative to the disk (as well as the triaxiality of the halo) can vary significantly over time. Therefore, the evolution of the disk warp is more plausibly governed by changes in the tilt angle, such as the variation $\Delta \theta$.

  We extended our analysis to all inclination groups and identified a clear dichotomy between short-term oscillations and long-term secular evolution. On short timescales (where the halo tilt fluctuates by $1^{\circ}\text{--}2^{\circ}$), the anti-correlation between the acceleration amplitude on the $Z=0$ plane and the warp amplitude is pronounced, confirming the angular momentum exchange (`seesaw') mechanism. However, this anti-correlation becomes difficult to capture over long timescales, where it is masked by the global alignment of the system. Specifically, in high-inclination models (e.g., $i=75^{\circ}$ inclination during $3\text{--}11$ Gyr, $\theta_{tilt}$ from $\sim38^\circ$ to $\sim0^\circ$; $i=60^{\circ}$ during $4.5\text{--}7.5$ Gyr, $\theta_{tilt}$ from $\sim35^\circ$ to $\sim10^\circ$; and $i=55^{\circ}$ during $5.5\text{--}9$ Gyr, $\theta_{tilt}$ from $\sim30^\circ$ to $\sim0^\circ$ as shown in Figure~\ref{F3}), both the halo tilt and the warp amplitude exhibit a simultaneous decline.
  
  Crucially, these periods of secular decay correspond to the phases of sustained prograde precession observed in Figure~\ref{F6}. As discussed in Section~\ref{sec:Kinematic warp model}, prograde precession ($\omega > 0$) is kinematically associated with a reduction in the disk's vertical velocity. This reduction reflects the continuous damping of the gravitational perturbation, indicating that the angular momenta of the DM halo and the disk are converging. This provides a physical explanation for why long-lived prograde precession is prevalent in high-inclination groups: the large initial misalignments ($\theta_{tilt}>20^{\circ}$) necessitate a prolonged phase of alignment. During this phase, the disk experiences a continuously decaying yet persistent gravitational torque, which sustains the prograde bending wave.

\section{Discussion}\label{sec:Discussion}
   \subsection{Comparing with the Galactic warp}\label{sec: cwtgw}
   
   In this work, we construct a low-mass model of the Milky Way that fits well with the observed rotation curve. Our goal is to investigate how the disk warp evolves under different orbital configurations in merger for the MW-like galaxies. Recent studies report that the Galactic warp amplitude reaches approximately  $\mathrm{1\,kpc}$ in $\mathrm{R=16\,kpc}$ \citep{Chen, He, Huang}, accompanied by a decreasing high prograde precession rate in the inner regions, followed by an increase at larger radii \citep{Dehnen23, Zhou}. We aimed to identify a simulation snapshot within the time interval $\mathrm{t=8\sim11\,Gyr}$ (corresponding to the epoch of the GSE merger) that matches the observational features. However, after examining all available snapshots, we did not find a satisfactory match.

   As discussed in the previous section, the primary source of disk disturbance in our model is the gravitational influence of the DM halo. In this merger model, the dynamic coupling between the DM halo and the disk in the later evolutionary stages, resulting in reduced gravitational disturbance. This trend is evident in Figure~\ref{F5} and the right panel of Figure~\ref{F11}. During the evolution, the disk warp undergoes repeated cycles of disappearance and regeneration. However, the peak amplitude of each regeneration event gradually decreases over time, which agrees with the observed general decrease in warp amplitude with decreasing redshift \citep{Vladimir}. Since gravitational disturbance scales with mass, this low-mass model struggles to produce high warp amplitudes in the later stages. As noted in Section~\ref{sec:Kinematic warp model}, observed prograde precession tends to reduce vertical velocity and thereby suppress warp amplitude. This suggests that the Galactic warp may have had a larger amplitude in the past. Therefore, an additional source of gravitational disturbance may be required to fully reproduce the Galactic warp. Previous studies suggest that the Sagittarius (Sgr) dwarf galaxy and the Large Magellanic Cloud (LMC) could also induce disk warp \citep{Stelea}.

   Some studies also report retrograde precession in observations \citep{Huang}, or more complex precession patterns that vary between low prograde and retrograde rates \citep{PH}. Such states may be more readily captured in simulations. Current observations also suggest a low precession rate of approximately $4.86\pm(0.88)_{stat}\pm(2.14)_{sys}$ km s$^{-1}$ kpc$^{-1}$ beyond 12.5 kpc \citep{Zhou25}. This value is significantly lower than the high prograde rates associated with rapid damping phases in our simulations (e.g., the $3\text{--}3.5$ Gyr and $5\text{--}5.5$ Gyr intervals in the $i=15^{\circ}$ model). This suggests that the Galactic warp is not currently undergoing a rapid collapse driven by a strong restoring torque. Instead, it implies that the Milky Way is in a stage of slow attenuation, where the halo and disk are gradually aligning. This scenario is consistent with a quiescent recent accretion history, where the warp is driven by a gentle, long-term interaction with a tilted DM halo.

   \subsection{Comparing with the baryonic vertical acceleration}\label{sec: thic}
   
   \begin{figure*}
   \label{F13}
   \centering
      \includegraphics[scale = 0.5]{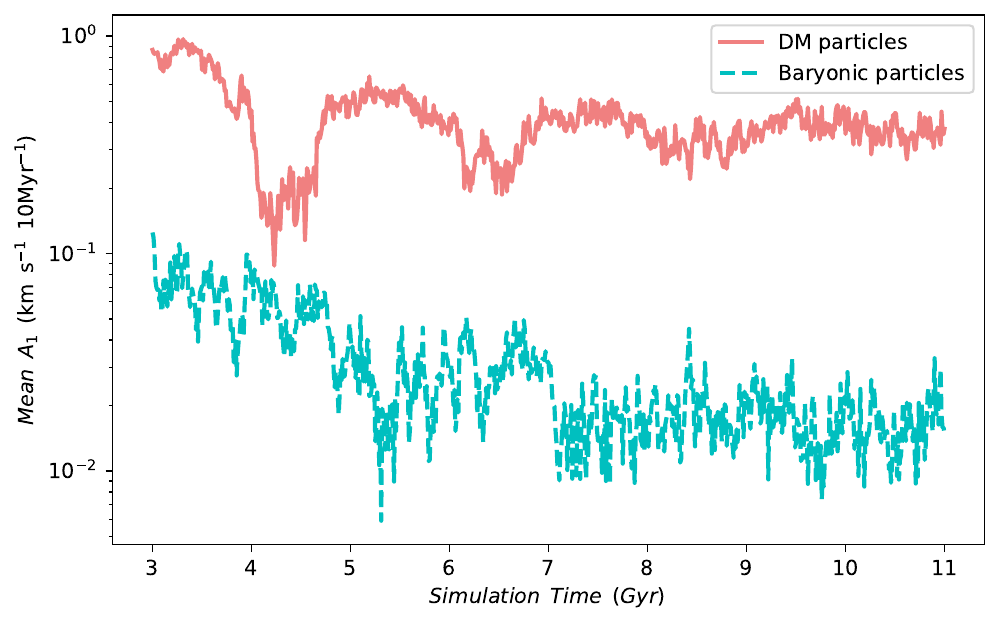}
  \caption{Comparison of the time evolution of the mean $m=1$ vertical acceleration amplitude $A_1$ acting on the $Z=0$ plane, generated by different galactic components (in logarithmic scale). The orange solid line represents the acceleration contribution from the dark matter halo, while the cyan dashed line shows the contribution from external baryonic structures. Although the baryonic components constitute a non-negligible mass fraction ($\sim 27\%$ of the enclosed DM mass within 20 kpc), their resulting vertical acceleration is consistently weak—generally less than one-tenth of the halo's intensity, except during the halo's local minimum at $4\text{--}5$ Gyr.}
         \label{F13}
   \end{figure*}

   In this subsection, we investigate whether external baryonic structures contribute significantly to driving the disk warp. Ignoring the self-gravity of the disk, we isolate the contribution from other baryonic components (corresponding to GADGET particle types 2 and 3 in the simulation). The total mass of these components is approximately $\mathrm{3.5\times10^{10}\,M_{\odot}}$  which corresponds to roughly 27\% of the dark matter halo mass enclosed within $\mathrm{R\leq20\,kpc}\,(\sim1.3\times10^{11}\,M_{\odot})$.

   Using the same methodology described in Section~\ref{sec: Perturbation from the DM halo}, we calculated the vertical gravitational acceleration $a_{Z}$, acting on the $Z=0$ plane and derived the time evolution of its $m=1$ mode amplitude. The results are presented in Figure~\ref{F13}.The orange solid line illustrates the vertical acceleration amplitude generated by dark matter particles, while the cyan dashed line shows that from the baryonic particles.

   We find that the gravitational intensity from the dark matter halo exhibits a more pronounced evolutionary trend, characterized by distinct peaks and troughs, which reflects the time-varying geometric tilt of the halo relative to the disk (as discussed in Section~\ref{sec: Perturbation from the DM halo}). In contrast, the acceleration produced by baryonic matter displays larger scatter on short timescales. After an initial decline, it stabilizes at a relatively low intensity. Quantitatively, the acceleration driven by baryonic particles is generally less than one-tenth of that provided by dark matter particles, except for the interval between 4–5 Gyr when the halo contribution is at a local minimum. Furthermore, we find no obvious correlation between the baryonic acceleration and the evolution of the warp amplitude. Consequently, we conclude that, unlike the dark matter halo, the external baryonic mass is not the primary driver of the disk warp.

   \subsection{Comparing with the flatten DM halo model}\label{sec: flatten DM halo}
   
   \begin{figure*}
   \label{F14}
   \centering
      \includegraphics[scale = 0.5]{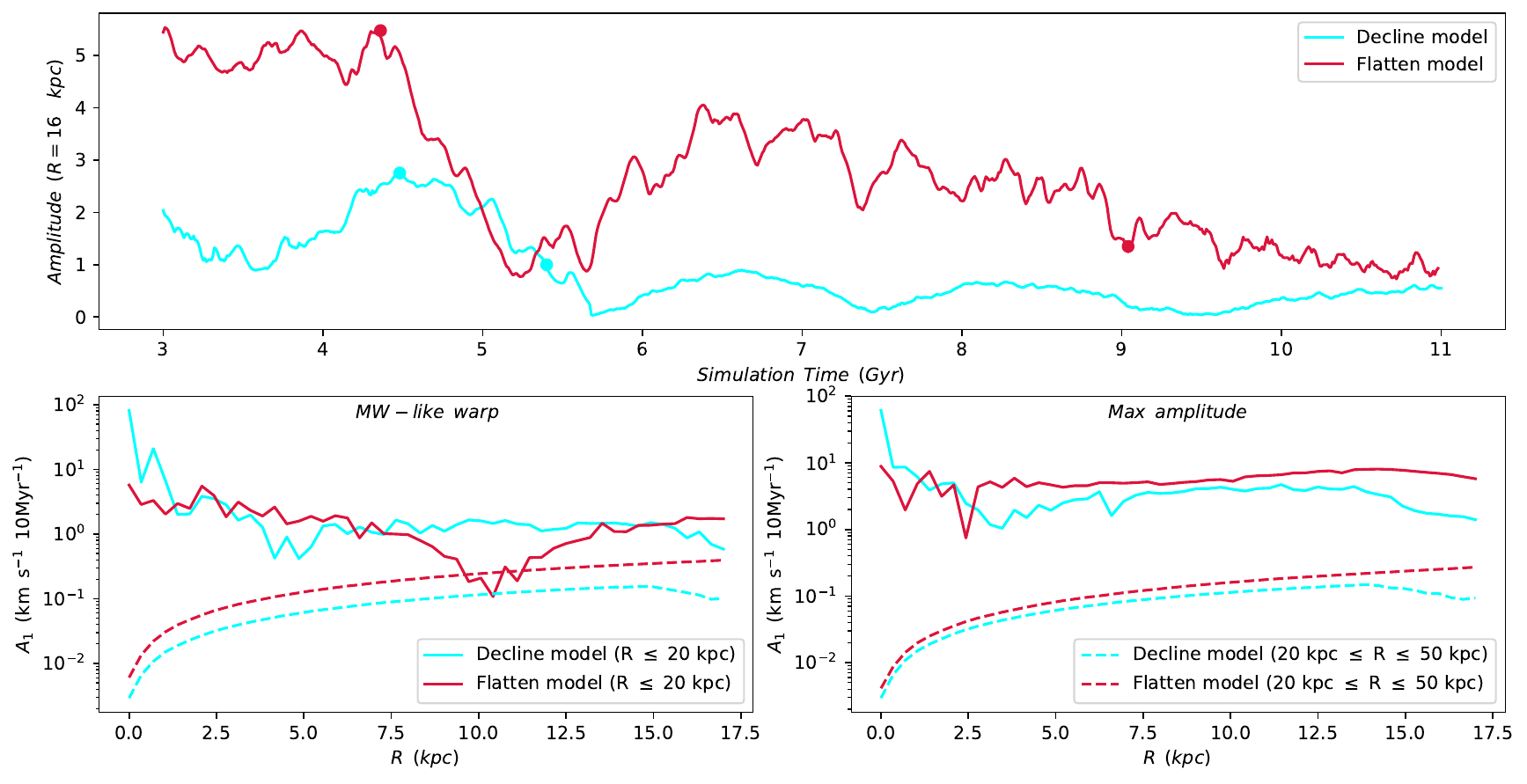}
  \caption{Comparison of warp generation under different Dark Matter halo potentials: the standard "flattened" halo model (Hernquist profile) and the "declining" halo model (Einasto profile). $Top\enspace panel$ shows the time evolution of the disk warp amplitude for both models. The colored dots mark two characteristic epochs for analysis: the state matching the Milky Way's current warp amplitude ("MW-like") and the state of maximum warp amplitude.
  $Bottom\enspace panels$ are the radial profiles of the $m=1$ vertical acceleration amplitude on the disk plane, analyzed at the "MW-like" epoch ($left$) and the "Max Amplitude" epoch ($right$). The solid lines represent the acceleration contribution from the inner halo ($R \le 20$ kpc), while the dashed lines represent the contribution from the outer halo ($20 \le R \le 50$ kpc). The flattened model produces a stronger maximum warp $\sim 5.4$ kpc compared to the Declining model $\sim 2.7$ kpc. Crucially, the decomposition reveals that the warp is predominantly driven by the inner halo ($R \le 20$ kpc, solid lines), where the acceleration is 1–2 orders of magnitude stronger than that from the outer halo (dashed lines).}
         \label{F14}
   \end{figure*}

   In our previous study, we successfully reproduced the disk warp using a gas-rich GSE merger model, in which the progenitor galaxy was modeled following \citet{Naidu2022}. In that model, the DM halo was represented using a Hernquist profile \citep{Hernquist} and a standard flattened distribution. In the current work, we adopt an Einasto profile to reproduce the declining rotation curve, which lead to a different gravitational potential on the disk plane. Figure~\ref{F14} shows comparison of two different model's evolution of disk warp amplitude, which are both on a $55^\circ$ inclination, clearly indicating that the flattened DM halo model produces a stronger warp. We mark the moments when each model reaches its maximum amplitude and the MW-like warp amplitude with dots in the top panel of Figure~\ref{F14}. The bottom panels compare the $m=1$ mode acceleration amplitudes on the disk plane. The left panel shows the MW-like warp, with an amplitude of approximately $\mathrm{\sim1\,kpc}$ at $\mathrm{R=16\,kpc}$), where solid lines represent the DM acceleration within $\mathrm{R\leq20\,kpc}$ and the DM mass for decline model is about $\mathrm{1.3\times10^{11}\,M_{\odot}}$, for the flatten model is $\mathrm{1.9\times10^{11}\,M_{\odot}}$, and dashed lines show the acceleration from $\mathrm{20\,kpc\leq R\leq 50\,kpc}$, the DM mass in this range is about $\mathrm{3.3\times10^{10}\,M_{\odot}}$ and $\mathrm{2.24\times10^{11}\,M_{\odot}}$ for decline and flatten model respectively. We observe that similar warp amplitudes can result from similar acceleration magnitudes, even in different mass distribution models. In the declining DM model, most of the mass is concentrated in the inner region ($\mathrm{R\leq2.5\,kpc}$), resulting in gravitational acceleration amplitudes that are an order of magnitude higher than those of the flattened model in the inner region. Specifically, the outer halo acceleration is one to two orders of magnitude weaker than that from the central region. The right panel shows the maximum warp amplitudes for each model: approximately $\mathrm{\sim5.4\,kpc}$ for the flattened model and $\mathrm{\sim2.7\,kpc}$ for the declining model. While the outer halo acceleration remains nearly unchanged, the central DM halo shows a significant increase in acceleration. The central acceleration amplitude in the flattened model is roughly twice that of the declining model.

   As shown in Figure~\ref{F14}, although the models adopt different DM halo profiles, the resulting disk warps are driven by similar gravitational strengths originating from the inner halo. An increase in acceleration on the disk plane corresponds to an increase in disk warp amplitude.

   Although the decline model may not be sufficient on its own to fully explain the evolution of the MW warp during its later stages, its contribution should not be overlooked. The model may still account for part of the present-day warp amplitude. Another scenario is proposed by \citet{Koop}, who suggest that repeated perturbations from satellite passages have significantly disturbed the MW disk, rendering its rotation curve unreliable for mass estimation. This might imply that the halo profile is still flattened in reality.

\section{Conclusion}\label{conclusion}

   In this work, we constructed a new model to reproduce the Galactic rotation curve by reconstructing the GSE merger. We varied the orbital inclination angles using the same galaxy progenitors to investigate how the disk warp evolves after the major merger. We found that higher inclination angles lead to stronger disk warps in the early stages of evolution than the lower inclination group. The warp follows a quasi-periodic pattern: it grows over time, reaches a peak, then gradually declines. Moreover, the peak amplitude decreases with each cycle and gradually reaches a stable state. Simultaneously, the warp undergoes precession: long-lasting low prograde precession typically appears in high-inclination cases, while retrograde precession is more common in low-inclination cases. Notably, strong prograde precession is typically observed during phases of rapidly declining warp amplitude. We further explored the mechanism behind disk warping, identifying the asymmetric gravitational potential of the DM halo as a key trigger. The morphological feature known as the tilted angle is a critical factor in quantifying this asymmetry. However, it reflects the gravitational potential in the $\mathrm{Z=0}$ plane, not necessarily on the actual disk plane. As the tilted angle increases, the gravitational acceleration on the $\mathrm{Z=0}$ plane also increases, and vice versa. Their evolution trends are consistent. Building on this analysis, we conclude that the interaction between the tilted halo and the disk warp operates in two distinct way. On short timescales involving small angular fluctuations, the system is governed by angular momentum exchange, manifesting as an anti-correlation between the driving halo torque and the disk's warp amplitude. Conversely, on secular timescales, the angular momenta of the halo and disk gradually tend toward alignment, leading to a simultaneous damping of both the halo tilt and the warp strength. This evolutionary framework also elucidates the prevalence of sustained prograde precession in high-inclination mergers: stemming from the high initial inclination between the dark matter halo and the disk plane, the disk is subjected to a gravitational perturbation that, while continuously decaying due to global alignment, remains persistent enough to sustain the prograde bending mode over extended periods. However, this low-mass decline halo model fails to reproduce the strong prograde precession observed in the MW. Future work may address this limitation by incorporating additional perturbers such as the Sgr and the LMC.

\begin{acknowledgements}

   This work was supported by National Key R\&D Program of China No. 2024YFA1611900, and the National Natural Science Foundation of China (NSFC Nos. 11973042, 11973052). We are grateful to Phil Hopkins and Jianling Wang for sharing access to the Gizmo code.
   
\end{acknowledgements}


\begin{thebibliography}{}
  \bibitem[Bailin(2003)]{Bailin}Bailin J. 2003, ApJL, 583, L79
  \bibitem[Barnes(2002)]{Barnes}Barnes J. E., 2002, MNRAS, 333, 481
  \bibitem[Battaner \& Jiménez-Vicente(1998)]{BJ}Battaner, E., \& Jiménez-Vicente, J. 1998, A\&A, 332, 809
  \bibitem[Belokurov et al.(2018)]{Belokurov} Belokurov V., Erkal D., Evans N. W., et al. 2018, MNRAS, 478, 611
  \bibitem[Belokurov et al.(2023)]{Belokurov23}Belokurov V., Vasiliev E., Deason A. J., et al. 2023, MNRAS, 518, 6200
  \bibitem[Bennett et al.(2022)]{Bennett}Bennett, M., Bovy, J., \& Hunt, J. A. S. 2022, ApJ, 927, 131
  \bibitem[Binney \& Tremaine(2008)]{BT}Binney J., Tremaine S. 2008, Galactic Dynamics, 2nd edn. Princeton Univ. Press, Princeton, NJ

  \bibitem[Bignone et al.(2019)]{Bignone}Bignone L. A., Helmi A., Tissera P. B. 2019, ApJL, 883, L5
  \bibitem[Bosma(1991)]{Bosma}Bosma A. 1991, Warped and Flaring HI Disks 
  (Cambridge: Cambridge Univ. Press), 181
  \bibitem[Cabrera-Gadea et al.(2024)]{CC}Cabrera-Gadea M., Mateu C., Ramos P., et al. 2024, MNRAS, 528, 4409
  \bibitem[Gaia Collaboration et al.(2023)]{G3} Gaia Collaboration et al. 2023b, A\&A, 674, A1
  \bibitem[Ciucă et al.(2024)]{Ciuca}Ciucă I., Kawata D., Ting Y.-S., et al. 2024, MNRAS.528L.122C
  \bibitem[Chen et al.(2019)]{Chen}Chen X., Wang S., Deng L., et al. 2019, Nature Astronomy, 3, 320
  \bibitem[Cheng et al.(2020)]{Cheng}Cheng X., Anguiano B., Majewski S. R., et al. 2020, ApJ, 905, 49
  \bibitem[Coe(2010)]{Coe}Coe D. 2010, preprint (arXiv:1005.0411)
  \bibitem[Conroy et al.(2019a)]{Conroy}Conroy C., Bonaca A., Cargile P., et al. 2019a, ApJ, 883, 107
  \bibitem[Cox et al.(2008)]{Cox}Cox T. J., Jonsson P., Somerville R. S., Primack J. R., Dekel A. 2008, MNRAS, 384, 386
  \bibitem[Dehnen et al.(2023)]{Dehnen23}Dehnen W., Semczuk M., \& Schönrich R. 2023, MNRAS, 523, 1556
  \bibitem[Deng et al.(2024)]{Deng}Deng M. J., Du C. H., Yang Y.B. 2024, ApJ, 975, 28D
  \bibitem[Einasto(1965)]{Einasto}Einasto J. 1965, Trudy Inst. Astroz. Alma-Ata, 51, 87 
  \bibitem[Elias et al.(2020)]{Elias}Elias L. M., Sales L. V., Helmi A., Hernquist L. 2020, MNRAS, 495, 29
  \bibitem[Emami et al.(2021)]{Emami}Emami R., Genel S., Hernquist L., et al. 2021, ApJ, 913, 36
  

  \bibitem[Fattahi et al.(2019)]{Fattahi}Fattahi A., Belokurov V., Deason A. J., et al. 2019, MNRAS, 484, 4471
  \bibitem[Freudenreich et al.(1994)]{Freudenreich}Freudenreich H. T., Berriman G. B., Dwek E., et al. 1994, ApJL, 429, L69
  
  \bibitem[Gaia Collaboration et al.(2023a)]{Gaia3}Gaia Collaboration, Drimmel R., Romero-G´omez M., et al. 2023a, A\&A, 674, A37
  \bibitem[Goyary \& Singh(2023)]{GS}Goyary S. S.,\& Singh H. S. 2023, MNRAS, 526, 5756G
  \bibitem[Grand et al.(2020)]{Grand}Grand R. J. J., Kawata D., Belokurov V., et al. 2020, MNRAS, 497, 1603
  \bibitem[Hammer et al.(2009)]{Hammer}Hammer F., Flores H., Puech M., Yang Y. B., et al. 2009, A\&A, 507, 1313
  \bibitem[Han et al.(2023)]{Han23}Han J. J., Conroy C., Hernquist L., 2023NatAs...7.1481H
  \bibitem[He(2023)]{He}He Z. H. 2023, ApJ, 954L, 9H
  \bibitem[Helmi et al.(2018)]{Helmi}Helmi A., Babusiaux C., Koppelman H. H., et al. 2018, Natur, 563, 85
  \bibitem[Hernquist(1990)]{Hernquist}Hernquist L. 1990, ApJ, 356, 359
  
  \bibitem[Hopkins et al.(2009)]{Hopkins09}Hopkins P. F. 2009, in Jogee S., Marinova I., Hao L., Blanc G. A., eds, ASP Conf. Ser. Vol. 419, Galaxy Evolution: Emerging Insights and Future Challenges. Astron. Soc. Pac., San Francisco, p. 228
  \bibitem[Hopkins(2015)]{Hopkins15}Hopkins P. F. 2015, MNRAS, 450, 53
  \bibitem[Hopkins et al.(2018)]{Hopkins18}Hopkins P. F., Wetzel A.,  Kereš D. 2018b, MNRAS, 480, 800
  \bibitem[Huang et al.(2024)]{Huang}Huang Y.,Feng Q., Khachaturyants T., et al. 2024, Natur, 8, 1294
  \bibitem[Hunter \& Toomre(1969)]{HT}Hunter, C. \& Toomre, A. 1969, ApJ, 155, 747
  \bibitem[Jiao et al.(2023)]{Jiao}Jiao Y. J., Hmmer F., Wang H. F., et al. 2023A\&A, 678A, 208J
  \bibitem[Jiang \& Binney(1999)]{JB}Jiang I.-G., \& Binney J. 1999, MNRAS, 303, L7
  \bibitem[Kazantzidis et al.(2004)]{Kazantzidis}Kazantzidis S., Kravtsov A. V., Zentner A. R., et al. 2004, ApJL, 611, L73
  \bibitem[Kerr(1957)]{Kerr}Kerr F. J. 1957, AJ, 62, 93
  \bibitem[Koop et al.(2024)]{Koop}Koop O., Antoja T., Helmi A., et al. 2024,
  A\&A, 692A, 50K
  \bibitem[Koppelman et al.(2020)]{Koppelman}Koppelman H. H., Bos R. O. Y., Helmi A. 2020, A\&A, 642, L18
  \bibitem[Lane et al.(2023)]{Lane}Lane J. M. M. , Bovy J., Mackereth J. T. 2023, MNRAS , 526, 1209 
  \bibitem[Laporte et al.(2018a)]{Laportea}Laporte C. F. P. , G´omez F. A., Besla G., et al. 2018, MNRAS , 473, 1218 
  \bibitem[Laporte et al.(2018b)]{Laporteb}Laporte, C. F. P., Johnston, K. V., G´omez, F. A.,
  et al. 2018, MNRAS, 481,286
  \bibitem[Lee et al.(2018)]{Lee}Lee, C. T., Primack, J. R., Behroozi, P., et al. 2018, MNRAS, 481, 4038
  \bibitem[Levine et al.(2006)]{Levine}Levine E. S., Blitz L., Heiles C. 2006, Science, 312, 1773
  
  \bibitem[López-Corredoira et al.(2002a)]{LC}López-Corredoira M., Betancort-Rijo J., \& Beckman J. E. 2002a, A\&A, 386, 169

 
  \bibitem[López-Corredoira et al.(2014)]{LC14}López-Corredoira M., Abedi H., Garzón F., et al. 2014, A\&A, 572, A101
  \bibitem[Li et al.(2023)]{Li23}Li X., Wang H.-F., Luo Y.-P., et al. 2023, ApJ, 943, 88
  \bibitem[Metropolis et al.(1953)]{Metropolis}Metropolis N., Rosenbluth A. W., Rosenbluth M. N., et al. 1953, J. Chem. Phys., 21, 1087
  \bibitem[Mowla et al.(2019)]{Mowla}Mowla L. A., van Dokkum P., Brammer G. B., et al. 2019, ApJ, 880, 57
  

  \bibitem[Naidu et al.(2020)]{Naidu2020}Naidu R. P., Conroy C., Bonaca A., et al. 2020, ApJ, 901, 48
  \bibitem[Naidu et al.(2021)]{Naidu2022}Naidu R. P., Conroy C., Bonaca A., et al. 2021, ApJ, 923, 92
  \bibitem[Nelson \& Tremaine(1995)]{NT}Nelson R. W. , Tremaine S. 1995, MNRAS , 275, 897 
  \bibitem[Ostriker \& Binney(1989)]{OB}Ostriker E. C., \& Binney J. J. 1989, MNRAS, 237, 785
  \bibitem[Ou et al.(2023)]{Ou}Ou X. W., Eilers A.C., Necib L., et al. 2024, MNRAS, 528, 693O
  \bibitem[Park et al.(2021)]{Park}Park M. J., Yi S. K., Peirani S., et al. 2021, ApJS, 254, 2
  \bibitem[Peng \& He(2024)]{PH}Peng L.M., He Z.H. 2024, arXiv:2412.20344
  \bibitem[Perret et al.(2014)]{Perret14}Perret V., Renaud F., Epinat B., et al. 2014, A\&A, 562, A1
  \bibitem[Perret(2016)]{Perret16}Perret V. 2016, DICE: Disk Initial Conditions Environment, Astrophysics Source Code Library, ascl:1607.002
  \bibitem[Poggio et al.(2020)]{Poggio}Poggio E., Drimmel R., Andrae R., et al. 2020, Nature Astronomy, 4, 590
  \bibitem[Poggio et al.(2021)]{Poggio21}Poggio, E., Laporte, C. F. P., Johnston, K. V., et al. 2021, MNRAS
  \bibitem[Pontzen et al.(2013)]{Pontzen}Pontzen A., Roškar R., Stinson G. S., et al. 2013, Astrophysics Source Code Library, record ascl:1305.002
  \bibitem[Quinn \& Binney(1992)]{QB}Quinn T., \& Binney J. 1992, MNRAS, 255, 729

  \bibitem[Revaz \& Pfenniger(2004)]{RP}Revaz Y., \& Pfenniger D. 2004, A\&A, 425, 67

  \bibitem[Reshetnikov \& Combes(1998)]{Reshetnikov}Reshetnikov V., \& Combes F. 1998, A\&A, 337, 9
  \bibitem[Rocha et al.(2008)]{Rocha}Rocha M., Jonsson P., Primack J. R., et al. 2008, MNRAS, 383, 1281
  \bibitem[Roškar et al.(2010)]{Roskar}Roškar R., Debattista V. P., Brooks A. M., et al. 2010, MNRAS, 408, 783
  \bibitem[Sánchez-Saavedra et al.(1990)]{SS90}Sánchez-Saavedra M. L., Battaner E., \& Florido E. 1990, MNRAS, 246, 458
  \bibitem[Sánchez-Saavedra et al.(2003)]{SS03}Sánchez-Saavedra M. L., Battaner E., Guijarro A., et al. 2003, A\&A, 399, 457
  \bibitem[Sellwood \& Debattista(2022)]{SD}Sellwood J. A., \& Debattista V. P. 2022, MNRAS, 510, 1375
  \bibitem[Semczuk et al.(2020)]{Semczuk}Semczuk, M., Łokas, E. L., D’Onghia, E., et al. 2020, MNRAS, 498, 3535
  \bibitem[Sérsic(1968)]{Sersic}Sérsic J. L. 1968, Atlas de galaxias australes (Cordoba, Argentina: Observatorio Astronomico, 1968)
  \bibitem[Shao et al.(2021)]{Shao}Shao S., Cautun M., Deason A., et al. 2021, MNRAS, 504, 6033
  \bibitem[Shen et al.(2006)]{Shen} Shen, J., \& Sellwood, J. A. 2006, MNRAS, 370, 2,
  \bibitem[Springel et al.(2005)]{Springel05}Springel V. 2005, MNRAS, 364, 1105
  \bibitem[Springel et al.(2021)]{Springel21}Springel V., Pakmor R., Zier O., et al. 2021, MNRAS, 506,2871
  \bibitem[Stelea et al.(2024)]{Stelea}Stelea I. A., Hunt J. A. S., \& Johnston K. V. 2024, ApJ, 977, 252S
  \bibitem[Vladimir et al.(2025)]{Vladimir}Vladimir P. R., Ilia V. C., Alexander A. M., et al.arXiv:2504.12403
  \bibitem[Wang et al.(2012)]{Wangj}Wang J., Hammer F., Athanassoula E., et al. 2012, A\&A, 538, A121
  \bibitem[Weinberg \& Blitz(2006)]{WB}Weinberg M. D., \& Blitz L. 2006, ApJL, 641, L33
  \bibitem[White \& Frenk(1991)]{White}White S. D. M., \& Frenk C. S. 1991, ApJ, 379, 52
  \bibitem[Xiang et al.(2025)]{Xiang}Xiang M., Rix HW., Yang H. et al, Nat Astron 9, 101–110.
  \bibitem[Zhou et al.(2024)]{Zhou} Zhou X., Chen X., Deng L., et al. 2024, ApJ, 965, 132.
  \bibitem[Zhou et al.(2025)]{Zhou25}Zhou X., Chen X., Deng L., et al. 2025, ApJ, 989, 213Z

\end{thebibliography}
\end{document}